\documentclass[journal]{IEEEtran}
\usepackage{graphicx}
\usepackage{enumerate}
\usepackage{subfigure}
\usepackage{balance}  % for  \balance command ON LAST PAGE  (only there!)
\usepackage{booktabs,threeparttable,caption}
\usepackage{verbatim}
\usepackage{bm}
\usepackage{stfloats}
\usepackage{amsmath,algorithmic,algorithm}
\usepackage{multirow,diagbox,makecell}
\usepackage[marginal]{footmisc}
\usepackage{amssymb}
\usepackage{lipsum}
\usepackage{multicol}

\newtheorem{definition}{Definition}

\begin{document}

\title{A Blockchain Transaction Graph based Machine Learning Method for Bitcoin Price Prediction}

\author{Xiao~Li and
	Weili~Wu
	
	\thanks{Corresponding Author: Xiao Li, Email: xiao.li@utdallas.edu}
	\thanks{X. Li and W. Wu are with the Department of Computer Science,
		The University of Texas at Dallas, Richardson, TX 75080 USA (e-mail:
		xiao.li@utdallas.edu; weiliwu@utdallas.edu).}
	\thanks{This work is in part supported by NSF grant 1747818 and 1907472.}
    }% <-this % stops a space

\markboth{}%
{Shell \MakeLowercase{\textit{et al.}}: Bare Demo of IEEEtran.cls for IEEE Journals}

\maketitle

\begin{abstract}
Bitcoin, as one of the most popular cryptocurrency, is recently attracting much attention of investors. Bitcoin price prediction task is consequently a rising academic topic for providing valuable insights and suggestions. Existing bitcoin prediction works mostly base on trivial feature engineering, that manually designs features or factors from multiple areas, including Bticoin Blockchain information, finance and social media sentiments. The feature engineering not only requires much human effort, but the effectiveness of the intuitively designed features can not be guaranteed. 
In this paper, we aim to mining the abundant patterns encoded in bitcoin transactions, and propose $k$-oder transaction graph to reveal patterns under different scope.  We propose the transaction graph based feature to automatically encode the patterns. A novel prediction method is proposed to accept the features and make price prediction, which can take advantage from particular patterns from different history period. The results of comparison experiments demonstrate that the proposed method outperforms the most recent state-of-art methods.
\end{abstract}


% \fi
\IEEEpeerreviewmaketitle

\section{Introduction}
\label{sec:intro}

Bitcoin Blockchain~\cite{Nakamoto2009BitcoinA}, as the very first application of blockchain, has been attracting more and more attention of public from various areas. The \emph{bitcoin} is the cryptocurrency traded in the Bitcoin blockchain, which is a reward to the miners for successfully mining a block. Bitcoin shares similar characteristics like many other financial products, eg. stocks, gold and crude oil~\cite{vassiliadis2017bitcoin}. The main characteristic of bitcoin is the volatility in price~\cite{Aalborg2019WhatCE,balcilar2017can}. Bitcoin was traded at 123.65 USD on September 30, 2013, and reached its peak at 18984.77 USD on December 18, 2017, then falls down at 9096.78 USD on July 3, 2020 \footnote{https://www.coindesk.com/price/bitcoin}. With the great opportunity of earning fortune from the striking difference in prices, bitcoin is becoming a popular financial asset, people make investment on it~\cite{Yermack2013IsB}.  Bitcoin's market capitalization in 2017 reached 300 billion US dollars, almost equal to that of Amazon in 2016~\cite{DBLP:journals/jcam/ChenLS20}.

Bitcoin price forecasting can provide valuable suggestions for investors to make decision that whether they can buy in more bitcoins to earn more, or it's better to sold their bitcoins out to avoid further shrink of assets. The basic bitcoin price forecasting only aims to predict the price trend, which only provide the suggestions on whether the price will rise or fall~\cite{DBLP:journals/jcam/ChenLS20, DBLP:conf/iscc/YaoXL19}.  The results of trend prediction can only present limited information to investors, investors always desire more accurate and more informative suggestions on the price, so that investors can make further analysis to evaluate how much will the price change impacts their assets. For example, if a trend prediction system suggests that the price will drop, investors may be panic and sold out all their bitcoins, however, if they can know that the price will only drop slightly, investors may choose to wait for further revival of price, which can avoid the loss of their asset. However, there is only handful work study the accurate price prediction to predict the price of bitcoin\cite{DBLP:conf/icdm/AbayAGKITT19, DBLP:conf/pakdd/AkcoraDGK18, DBLP:journals/asc/MallquiF19, DBLP:conf/www/CerdaR19}. Therefore in this paper, we study the accurate bitcoin price prediction that is much more critical and useful in practice.

To make the prediction of price, machine learning models, either classification models or regression models, are adopted as popular prediction models. Therefore well-designed features are required to feed the machine learning models. Existing work has proposed various features to encode the latent properties behind the Bitcoin price from multiple aspects.

The most basic features of blockchain are the indexes reflecting the information of blockchain, such as mean degree of addresses, number of new addresses, total coin amount transferred in transactions and so on~\cite{DBLP:conf/icdm/AbayAGKITT19}. 
Maesa et. al also analysis the features of blockchain by constructing a full users graph~\cite{DBLP:conf/dsaa/MaesaMR16}.
Mallqui et. al.~\cite{DBLP:journals/asc/MallquiF19} include international economic indicators to reflect the feature of the financial market, such as crude oil future prices, gold future prices, S\&P500 future, NASDAQ future, and DAX index, which are features from financial perspective. 
CerdaR et. al.~\cite{DBLP:conf/www/CerdaR19} introduces public opinion feature into bitcoin price prediction through mining the sentiment from social media like Twitter. Yao et. al~\cite{DBLP:conf/iscc/YaoXL19} attempt to represent the opinion feature from news articles. These public opinion features are considered based on the intuition that people will take action based on how much the positive or negative opinions delivered by news articles and social media. 

Those existing work mostly manually define features for the bitcoin price prediction tasks. Although the created features cover many aspects, including blockchain network, financial market information, and even public opinions, the feature engineering is trivial and there are also many latent features that are hard to define them explicitly and manually. On the other hands,  if the external factors that beyond the Bitcoin blockchain, such as public opinions or the financial market, contribute to the price change of bitcoin, they will eventually be reflected by the changes in the Bitcoin blockchain, because the external factors will influence the action of users, then the user's action will be reflected by the transactions in Bitcion Blockchain. In this paper, we argue that Bitcoin blockchain encodes abundant information that have latent pattern to represent the features behind the bitcoin price. 

We believe that the patterns of the transaction is very expressive that the patterns can directly reflect what's going on the  in the blockchain, and thus represents financial status of bitcoin market. As mentioned in~\cite{DBLP:conf/pakdd/AkcoraDGK18}, if a input addresses of a transaction is more than the output addresses, then the transaction is gathering bitcoins, indicating some users are buying bitcoins. On the other hand, if the input addresses of a transaction is less than the output addresses, then the transaction is splitting the bitcoins, indicating some users are selling bitcoins. Therefore the patterns of the transactions can represent very useful information that can hardly managed by manual feature engineering. 

To effectively mining the transaction patterns, we employ the transaction graph to represent the bitcoin blockchain. Further we propose the $k$-order transaction subgraph to encode the transaction pattern, with different $k$, different level of patterns. Finally the pattern occurrence matrix is propose to store the frequency of the patterns occurred in blockchain, which can represent the feature of a period of blockchain.

The main contributions of the paper can be summarized as follows:
\begin{itemize}
	\item We propose the $k$-order transaction subgraph based on the transaction graph, to represent the transaction feature of Bitcoin Blockchain. 
	\item We proposed transaction graph based feature to encode the latent patterns behind the transactions, which is further fed in a novel machine learning based prediction method that can effectively learn the  characteristics of different history period. 
	\item To our best knowledge, we are the first to utilize only Bitcoin Blockchain transaction information, to tackle the price prediction problem. 
	\item We evaluate proposed method on real bitcoin price historical data, and the results demonstrate the superiority comparing to recent state-of-the art methods.
\end{itemize}

The remainder of this paper is organized as follows: In Section~\ref{sec:rel_wo}, recent related work is presented. Then we describe our proposed transaction graph based price prediction methodology in Section~\ref{sec:meth}. In Section~\ref{sec:exp}, we evaluate the proposed method. Finally, the conclusion is made in Section~\ref{sec:con}.

\section{Related Work}
\label{sec:rel_wo}

The key issue of bitcoin price prediction or forecasting task is to discover and analysis determinants of bitcoin price.  Since Ladislav Kristoufek~\cite{kristoufek2013bitcoin} studied the connection between Bitcoin and search queries on Google Trends and Wikipedia, the determinants study has developed rapidly.  Kristoufek's results show a positive correlation between a price level of Bitcoin and the searched terms, and an asymmetry between effects of search queries related to prices above and below a short-term trend.

Extending from analyzing the relation of Wikipedia and Google search queries, researchers also evaluate the influence of social media or public opinions~\cite{DBLP:conf/www/CerdaR19, DBLP:conf/iscc/YaoXL19}. Balfagih and Keselj ~\cite{DBLP:conf/bigdataconf/BalfagihK19} extensively explored the relationship between bitcoin tweets and the prices, which utilizes different language modeling approaches, such as tweet embedding and N-Gram modeling. Polaski et. al ~\cite{DBLP:journals/ijecommerce/PolasikPWKL15} discover that the bitcoin prices are primarily driven by the popularity of Bitcoin, the sentiment expressed in newspaper reports on cryptocurrency, and total number of transactions. Mittal et. al~\cite{DBLP:conf/ic3/MittalDSP19} argues that there is a relevant degree of correlation of Google Trends and Tweet volume data with the price of Bitcoin, while no significant relation with the sentiments of tweets is discovered. Piwowarski et. al\cite{DBLP:conf/sigir/BurnieY19} aim at analyzing particular relevance of topics on social media for the high volatility of bitcoin price when shifting in distinct four phases across 2017 to 2018.

Georgoula et.al~\cite{DBLP:conf/mcis/GeorgoulaPBSG15} show that the sentiment ratio of twitter has positive correlation with bitcoin prices, while the value of Bitcoins is negatively affected by the exchange rate between the USD and the euro has negative relation with bitcoin price. Ciaian et. al~\cite{Ciaian16} are the first to studies BitCoin price formation by considering both traditional features in market and digital currencies specific factors from the economical aspects. Brandvold et. al~\cite{brandvold2015price} study the contributions of Bitcoin exchanges to price discovery, and demonstrate that Mt.Gox and BTC-e are the market leaders with the highest information share. Aggarwal et. al~\cite{DBLP:conf/ic3/AggarwalGGG19} attempt to compare the effects of  determinants including bitcoin factors, social media and the Gold price. Pieters and Vivanco~\cite{DBLP:journals/iepol/PietersV17} study the difference in Bitcoin prices across 11 different markets, and present that standard financial regulations can have a non-negligible impact on the market for Bitcoin. 

Both Georgoula et.al~\cite{DBLP:conf/mcis/GeorgoulaPBSG15} and Kristoufek~\cite{kristoufek2015main} studies the difference of long-term and short-term impact of the determinants on bitcoin price. Kristoufek~\cite{kristoufek2015main} stresses that both time and frequency are crucial factors for Bitcoin price dynamics since the bitcoin price evolves overtime, and examines how the interconnections from various sources behave in both time and different frequencies.  Bouoiyour et. al ~\cite{bouoiyour2016drives} adopt Empirical Mode Decomposition (EMD) for Bitcoin price analysis, and the results show that   the long-term fundamentals contributes most to itcoin price variation, though intuitively bitcoin market seems a short-term speculative bubble. Chen et. al  ~\cite{DBLP:journals/ijcse/ChenZMWZY20} analyze the dependence structure between price and its influence factors, and based on copula theory, the bitcoin price has different correlation structures with influence factors.

Bitcoin Blockchain structural information is also mined for discovering the features and determinants of the bitcoin prices. Akcora et. al ~\cite{DBLP:conf/pakdd/AkcoraDGK18} propose a bitcoin graph model, upon which Chainlets is proposed to represent  graph structures in the Bitcoin. A $k$-chainlet is a subgraph of a bitcoin model that contains exactly $k$ transaction nodes.  Akcora et. al ~\cite{DBLP:conf/pakdd/AkcoraDGK18} employ both the features derived from chainlets and heuristic features to fed in machine learning model for price prediction. In Akcora et. al's further work ~\cite{DBLP:conf/icdm/AbayAGKITT19}, they propose occurrence matrix and amount matrix to encode the topological features of chainlets. In this paper, we also adopt the concept of occurrence matrix to encode the topological features. However, we design totally different graph representation model to reveal the topological features of Bitcoin Blockchain.

The determinants can be considered as features behind the bitcoin price change, then various machine learning methods can be  adopted to learn the patterns from the features and make bitcoin price forecasting~\cite{DBLP:journals/jcam/ChenLS20, DBLP:conf/educon/YogeshwaranKM19, DBLP:conf/icnc/SinW17}. Felizardo et. al ~\cite{DBLP:conf/besc/FelizardoOHC19, DBLP:conf/ispacs/ChenCLHLW19} compare several popular machine learning methods adopted in bitcoin price prediction task, such as Back-propagation Neural Network (BPNN), Autoregressive Integrated Moving Average mode (ARIMA), Random Forest (RF), Support Vector Machine (SVM), Long Short-Term Memory (LSTM) and WaveNets. Wu et. al~\cite{DBLP:conf/icdm/WuLML18} proposed a novel LSTM with AR(2) model that outperforms conventional LSTM model. Hashish et. al~\cite{DBLP:conf/etfa/HashishFAFD19} use Hidden Markov Models to tackle the volatility of cryptocurrencies and predict the future movements with LSTM. Nguyen et. al~\cite{DBLP:conf/fdse/NguyenL19} propose hybrid methods between ARIMA and machine learning to predict the bitcoin next day price.

%	\begin{algorithm}[tb]
%		\caption{Signal Decision Algorithm }
%		\label{alg:SSF_Extraction}
%		\begin{algorithmic}[1]
%			\REQUIRE Location List:$ L$, Token list: $T$, current token: $t$, current location $l$.
%			\ENSURE warning signal: $w\in \{0,1\}$, alarm signal: $a \in \{0,1\}$
%			\STATE initialize $w \leftarrow 0$, $a \leftarrow 0$
%			\STATE $density \leftarrow \frac{length(L)}{\pi (0.005)^2} $
%			\IF {$density > N$}
%				\STATE $a \leftarrow 1$
%			\ELSE
%				\FOR {$i$ from 1 to $length(L)$ }
%					\IF {$T[i]$ == red}
%						\IF{distance($L[i],l$)<2}
%							\STATE $a \leftarrow 1$
%							\RETURN $w, a$
%						\ELSE
%							\STATE $w \leftarrow 1$
%						\ENDIF
%					\ENDIF
%				\ENDFOR		
%				\IF {$w ==0$ \& $density > M$}
%					\STATE $w \leftarrow 1$
%					\RETURN $w, a$
%				\ENDIF
%			\ENDIF
%			\RETURN $w, a$
%		\end{algorithmic}
%	\end{algorithm}

\section{Methodology}
\label{sec:meth}

\subsection{Problem Definition}
\begin{definition}
	\label{def:price_def}
	\emph{(Bitcoin Price Trend Prediction) :}
	Given timestamp $t+h$, where $h\in N^+$, and bitcoin historical data in time period $[t-i, t]$, where $i\in N^+$. Let $P_t$ denotes the price of bitcoin at the timestamp $t$. the bitcoin price trend prediction problem is to predict the label $L_{t+h} \in \{-1,1\}$, where $L_{t+h} =1$ if $P_{t+h}>P_t$, and $L_{t+h} =-1$ otherwise.
\end{definition}

Base on above basic Bitcoin Price Trend Prediction task, we define and study the Bitcoin Price Prediction task, which is defined in Definition~\ref{def:BPP}.

\begin{definition}
	\emph{(Bitcoin Price  Prediction) :}
	\label{def:BPP}
	Given timestamp $t+h$, where $h\in N^+$, and bitcoin historical data in time period $[t-i, t]$, where $i\in N^+$. Let $P_t$ denotes the price of bitcoin at the timestamp $t$. the bitcoin price prediction problem is to predict the price at timestamp $t+h$, $P_{t+h}$.
\end{definition}

To manage the Bitcoin Price Prediction task defined above, classic machine learning framework is employed in this paper. First, feature vectors are obtained from the Bitcoin Blockchain in the historical time period $[t-i, t]$. Then the feature vectors are fed in machine learning models to learn and predict the value of future bitcoin price. Since the values of prices is continuous, the price prediction problem can be considered as a classic regression problem.

\subsection{Method Overview}

\begin{figure*}[h]
	\centering
	\includegraphics[width=0.7\linewidth]{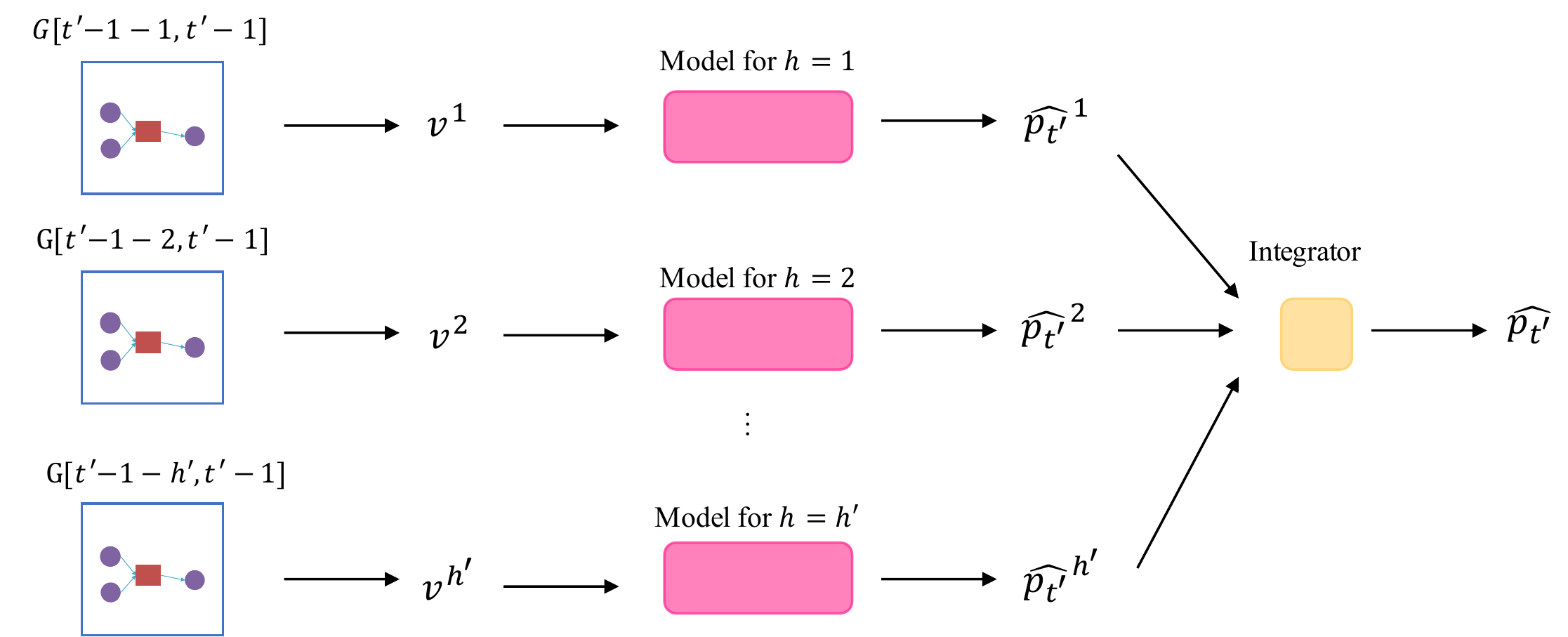}
	\caption{The overview of whole Method}
	\label{fig:framework}
\end{figure*}

Figure~\ref{fig:framework} illustrates an overview of the whole method. We propose to establish independent models for different $h$, which means each model only trained for predict specific future time. For example, the model for $h=1$ only trained for predicting the price at tomorrow, and model for $h=1$ is only for predicting the price at the day after tomorrow.  For predicting the bitcoin price at time $t'$, $h'$ independent models will be trained and produce $h'$ different values of the estimated price $\hat{p_{t'}}$, the integrator will then generate the final estimated value based on the output of all the independent models. 

Each independent machine learning model will take the features that generated from transaction graphs as inputs, and is trained separately. The features, e.g. $v^1, ...,  v^{h'}$ will be proposed in latter sections. Since the models are trained for different $h$, the time period utilized for the models should also be different. 
$h$ can also be considered as the length of the history window size, which is a hyper parameter to adjust how much historical information is considered in the the features. 
The start point and the end point of historical time period is adjusted based on the value of $h'$. There are totally $h'$ different periods utilized to learn the patterns to predict the price at the same day $\hat{p_{t'}}$.

The integrator can be a simple linear function, that: 

\begin{equation}
\hat{p_{t'}}  = \alpha_1 *\hat{p_{t'}}^1 + \alpha_2 *\hat{p_{t'}}^2 + ... + \alpha_{h'} *\hat{p_{t'}}^{h'}
\end{equation}

where $\alpha_1+ \alpha_2 + ... + \alpha_{h'} =1$. 

In this paper, we elaborately design the weights. Let $W_i =[\alpha_1, \alpha_2, \alpha_3, ..., \alpha_i]$. Specially, if the history window size is 1, which indicates we only employ one model to make prediction,$W_1 = [\alpha_1] = [1.0]$. With the increase of history window size, for $i>1$, the $W_i$ is defined as Equation~\ref{eq:w}: 
\begin{equation}
\label{eq:w}
\begin{aligned}	
W_{i+1}[k] & = W_i[k] \ (k=1,..., i-1) \\
W_{i+1}[i] &= W_i[i]*r \\
W_{i+1}[i+1] & = W_i[i]*(1-r)	 
\end{aligned}
\end{equation}
where $r$ controls the speed of decay of weights corresponding for further history. 
Equation~\ref{eq:w} maintains the property that $\sum_{\alpha_j \in W_i} \alpha_j = 1$ for $i>0$.

\subsection{Transaction Graph}
%Existing works mostly manually define the features based on domain knowledge for the prediction. Although the transparently defined feature are easy to interpret, manually defining the feature is not possible to mining the latent information in the bitcoin blockchain.
 In order to mining the bitcoin blockchain information, we obtain the feature by employ \textbf{\emph{transaction graph}} to represent the bitcoin blockchain.

There are similar concepts of transaction graph in literature\cite{DBLP:conf/icdm/AbayAGKITT19,DBLP:conf/dsaa/MaesaMR16}, here we define the transaction graph as following.

\begin{definition}
	\emph{(Transaction Graph) :}
	A \emph{transaction graph} is a directed graph $G = (A,T,E)$, where $A$ is the set of addresses in Blockchain, $T$ is the set of transactions of blockchain, and $E$ is the set of direct link from $a_i \in A$ to $t_k \in T$, indicating $a_i$ is one of the inputs of $t_k$, or from $t_k \in T$ to $a_j \in A$, indicating $a_j$ is one of the outputs of $t_k$.
\end{definition}

\begin{figure}[h]
	\centering
	\includegraphics[width=0.7\linewidth]{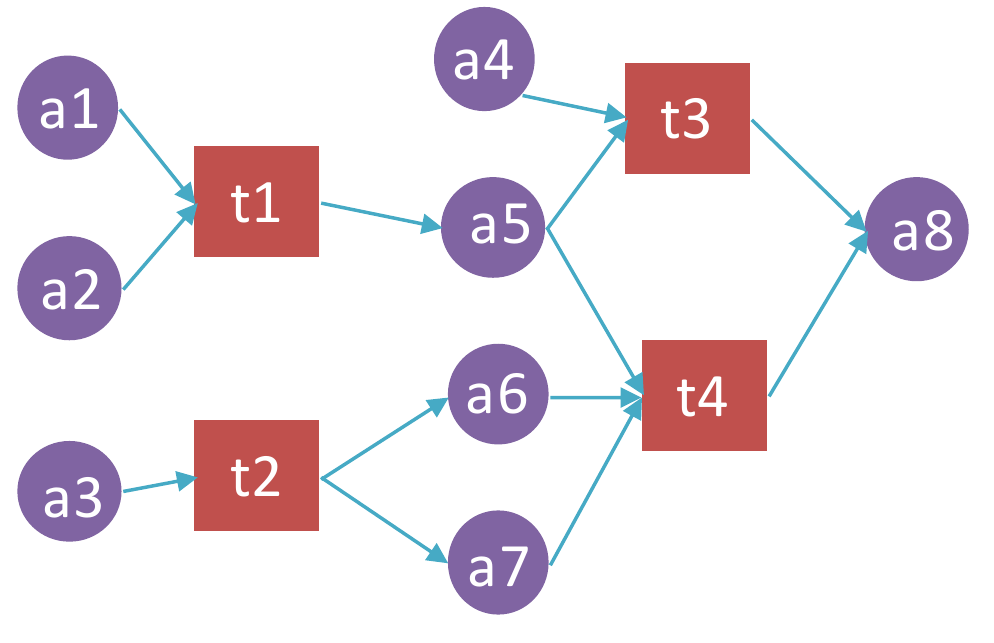}
	\caption{A Simple Transaction Graph}
	\label{fig:tx_graph}
\end{figure}

Figure~\ref{fig:tx_graph} presents a simple example of a transaction graph, which contains 8 addresses and 4 transactions. 

\subsection{k-order transaction subgraph}
The $k$-order transaction subgraph of a transaction $t_i$ is a graph $G_{t_i}^k$ that contains only $t_i$ and the transactions that spent the output of $t_i$ in next $k-1$ steps, and the corresponding addresses connecting to these transactions. The formal definition is given in 
Definition~\ref{def:k-order}.

\begin{definition}
	\label{def:k-order}
	\emph{($K$-order transaction subgraph) :}
	The $K$-order transaction subgraph of a transaction $t_i$ is a graph $G_{t_i}^k = (A^k,T^k,E^k)$, where $T^k= \{t_j|\ \exists \  a_n \in A^{k-1}, \ (a_n,t_j) \in E   \ and\  \exists (t_l,a_n)\in E^{k-1} \ for\ t_l \in T^{k-1}\}$, $A^k = \{a_n| a_n \in A^{k-1}  \ or\  (t_j,a_n)\in E \ where \ t_j \in T^k\}$. Specially, if $k=1$, $G_{t_i}^1 = (A^1, T^1,E^1)$, where $A^1 = \{a_n| (a_n,t_i)\in E \ or \ (t_i,a_n)\in E \}$, $T^1 = \{t_i\}$ and $E^1 = \{(a_n,t_i) \ or \ (t_i,a_n)| a_n\in A^1\}$.
\end{definition}

\begin{figure}[t]
	\centering
	\small	
	\subfigure[$1$-order transaction subgrpah of $t_1$, $G_{t_1}^1$]{
		\label{fig:1-order}
		\begin{minipage}[t]{0.55\linewidth}
			\centering
			\includegraphics[width=0.7\linewidth]{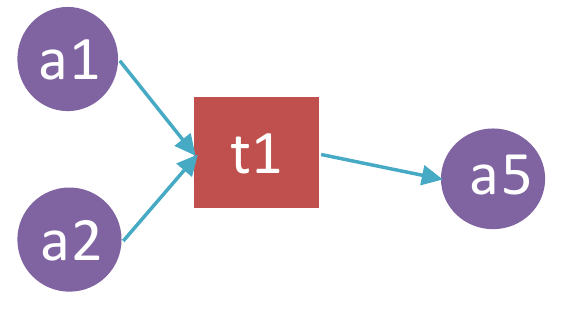}
		\end{minipage}
	}
	\hspace{0.1in}
	\subfigure[$1$-order transaction subgrpah of $t_2$, $G_{t_2}^1$]{
		\label{fig:1-order2}
		\begin{minipage}[t]{0.55\linewidth}
			\centering
			\includegraphics[width=0.7\linewidth]{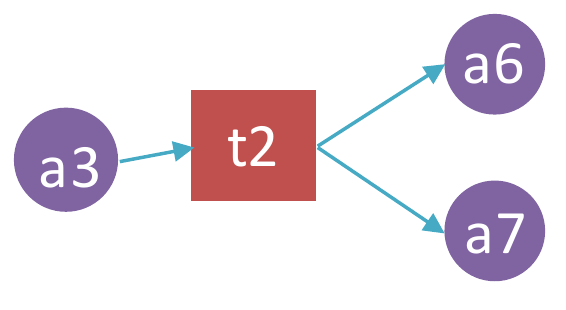}
		\end{minipage}
	}
	\hspace{0.1in}
	\subfigure[$2$-order transaction subgrpah of $t_1$, $G_{t_1}^2$]{
		\label{fig:2-order}
		\begin{minipage}[t]{1\linewidth}
			\centering
			\includegraphics[width=0.7\linewidth]{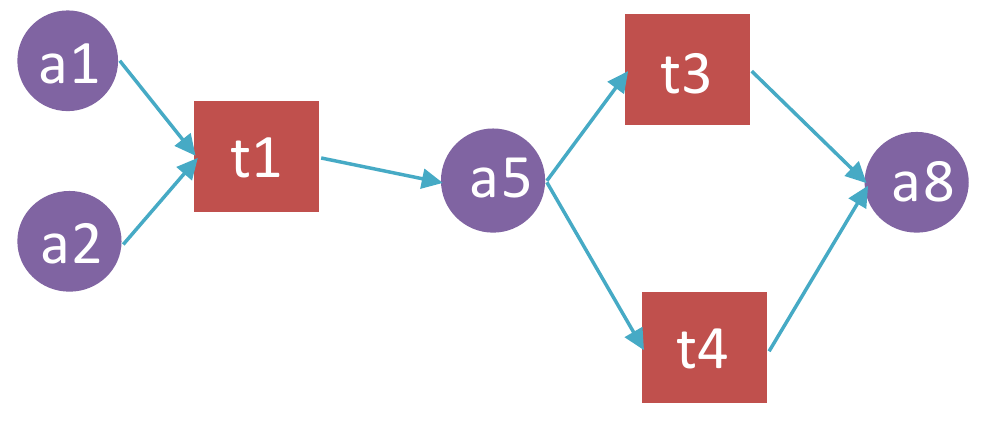}
		\end{minipage}
	}
	\hspace{0.1in}
	\subfigure[$2$-order transaction subgrpah of $t_2$, $G_{t_2}^2$]{
		\label{fig:2-order2}
		\begin{minipage}[t]{1\linewidth}
			\centering
			\includegraphics[width=0.7\linewidth]{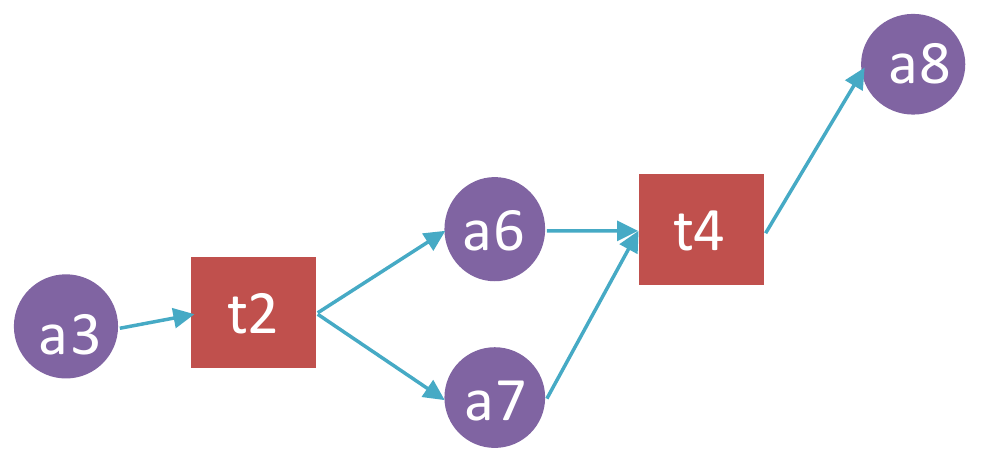}
		\end{minipage}
	}
	\caption{The $1$ order nd $2$-order transaction subgraph of $t_1$ in Figure~\ref{fig:tx_graph}}	
	\label{fig:k-sub}
\end{figure}

Obviously, if $k=1$, then the $K$-order transaction subgraph of $t_i$ contains only $t_i$ along with its input addresses and output addresses. With $k$ increases, the $k$ order transaction subgraph will trace further along with the bitcoin flow outputted by transaction $t_i$. Figure~\ref{fig:1-order} and ~\ref{fig:2-order} shows the $1$-order and $2$-order transaction subgraph of the transaction $t_1$ in Figure~\ref{fig:tx_graph}, respectively. Figure~\ref{fig:1-order2} and ~\ref{fig:2-order2} shows the $1$-order and $2$-order transaction subgraph of $t_2$, respectively.

The $k$-order transaction subgraphs have different pattern by considering the different structures of them. Here we consider different patterns as different numbers of inputs and outputs addresses of the $k$-order transaction subgraphs.

The input addresses of a $k$-order transaction subgraph $G_{t_i}^k$ is the addresses that only give inputs to transactions in $G_{t_i}^k$, and meanwhile do not accept any output from any transactions in $G_{t_i}^k$. Contrarily, the output addresses of  $G_{t_i}^k$ is the addresses that only accepts outputs of transactions in $ G_{t_i}^k$, and do not give any input from any transactions in  $G_{t_i}^k$. For clear notations, the definition of input and output addresses are defined in Definition~\ref{def:IO_k-order}.

\begin{definition}
	\label{def:IO_k-order}
	\emph{(Input and Output addresses of $K$-order transaction subgraph) :}
	The input and output addresses of $K$-order transaction subgraph $G_{t_i}^k$ is $\mathcal{I}_{G_{t_i}^k}$ and $\mathcal{O}_{G_{t_i}^k}$, respectively. $\mathcal{I}_{G_{t_i}^k} = \{ a_n| \exists (a_n, t_j) \in E^k, t_j \in T^k \ and\ \forall t_k \in T^k, (t_k,a_n) \notin E^k \}$. $\mathcal{O}_{G_{t_i}^k} = \{ a_n| \exists (t_k, a_n) \in E^k, t_k \in T^k \ and\ \forall t_j \in T^k, (a_n,t_j) \notin E^k \}$.
\end{definition}

In Figure~\ref{fig:1-order}, the addresses $a_1$ and $a_2$ are the input addresses of $G_{t_1}^1$, and the address $a_5$ is the output address of  $G_{t_1}^1$. For higher orders of  transaction subgraph, the input and output addresses may be more complicated. For example, in Figure~\ref{fig:2-order},  the input addresses of $G_{t_1}^2$ are $\{a_1, a_2\}= \mathcal{I}_{G_{t_1}^2}$, and the output addresses are $\{a_8\}=\mathcal{O}_{G_{t_1}^2}$. Please note that $a_5$ is not an input nor an output address, the function of $a_5$ in $G_{t_1}^2$ is only for transition of Bitcoins. Similarly, in Figure~\ref{fig:2-order2}, the input addresses of $G_{t_2}^2$ are $\{a_3 \} = \mathcal{I}_{G_{t_2}^2}$, while the output addresses of $G_{t_2}^2$ is $\{a_8\} = \mathcal{O}_{G_{t_2}^2}$.

Based on the concept of $\mathcal{I}_{G_{t_i}^k}$ and $\mathcal{O}_{G_{t_i}^k}$, we can further define the \textbf{\emph{pattern}} of a transaction subgraph. The pattern of a $k$-order transaction graph is denoted as $G_{(m,n)}^k = \{G_{t_i}^k| |\mathcal{I}_{G_{t_i}^k}|=m,  |\mathcal{O}_{G_{t_i}^k}| = n \}$, where $m$ and $n$ are the number of input addresses and output addresses of $G_{t_i}^k$ respectively.

For a given transaction graph generated from the transaction information in bitcoin blockchain during a specific period, we can obtain $k$ order transaction subgraph $G_{t_i}^k$ of each transaction $t_i\in T$. The obtained transaction subgraphs may belongs to different patterns. For the example in Figure~\ref{fig:k-sub}, $G_{t_1}^2$ belongs to the pattern $G_{(2,1)}^2$, while $G_{t_2}^2$ belongs to the pattern $G_{(1,1)}^2$.

We believe these patterns contain valuable information revealing the characteristics of the transaction graph of corresponding blockchain in a period.  Besides, the patterns obtained under different order $k$ can reveal different level of latent information. The benefit of denoting the pattern based on the number of input addresses and out addresses is that the patters can be easily encoded into matrices, and therefore can be adopted as the features  of the current transaction graph.

The two key information that can represent the features of a transaction graph are 1)what kinds of patterns occurred in the transaction graph, and 2) how many times of these patterns occurred. We extend the concept of occurrence matrix in literature~\cite{DBLP:conf/icdm/AbayAGKITT19} to $k$ order occurrence matrix, denoted as $OC^k$, where the entry of $OC^k$ is $OC^k_{(m,n)} = |G_{(m,n)}^k|$.

Finally we concatenate $OC^k$ for $k = 1,2,3, ..,s$ as the feature of the transaction graph $G$ that representing the corresponding blockchain period. Now the \emph{Bitcoin Price Prediction} problem in Definition~\ref{def:BPP} can be specified in detail: use the feature vector $v$ that is a concatenation of $OC^k$, which is calculated on the transaction graph based on the bitcoin historical data in time period $[t-i,t]$, to predict the bitcoin price at timestamp $t+h$, $P_{t+h}$. In this paper, we apply a trick on this prediction task, that instead of directly fitting the relation of the feature vector $v$ with $P_{t+h}$, we fit $v$ with $P_{t+h}-P_t$. The reason is that the feature actually is represent the new transaction data in past period, not the cumulative transaction data, so predicting the price difference rather than the price itself is more reasonable. 

\subsection{Computation of Occurrence Matrix}
The above sections give a comprehensive interpretation of the occurrence matrix. In this section, we propose an iterative manner for realistic implementation by multiplying matrices to efficiently compute the occurrence matrix.

\begin{figure}[h]
	\centering
	\includegraphics[width=0.4\linewidth]{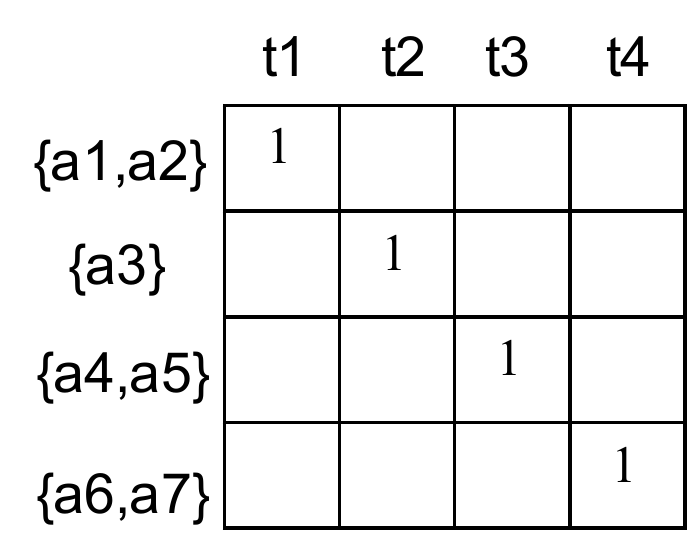}
	\caption{The $H$ matrix of Transaction Graph in Figure~\ref{fig:tx_graph}.}
	\label{fig:h_mat}
\end{figure}

Let $H \in \mathbb{R}^{|T|\times |T|}$ be the matrix denoting the input addresses of each transaction, The entry of $H$ is $H_{A_i,t_i}  = 1$, if $A_i$ is the set of input addresses of transaction $t_i$, otherwise $H_{A_i,t_i}  = 0$. Figure~\ref{fig:h_mat} shows the $H$ matrix of transaction graph in Figure~\ref{fig:tx_graph}.

\begin{figure}[h]
	\centering
	\includegraphics[width=0.4\linewidth]{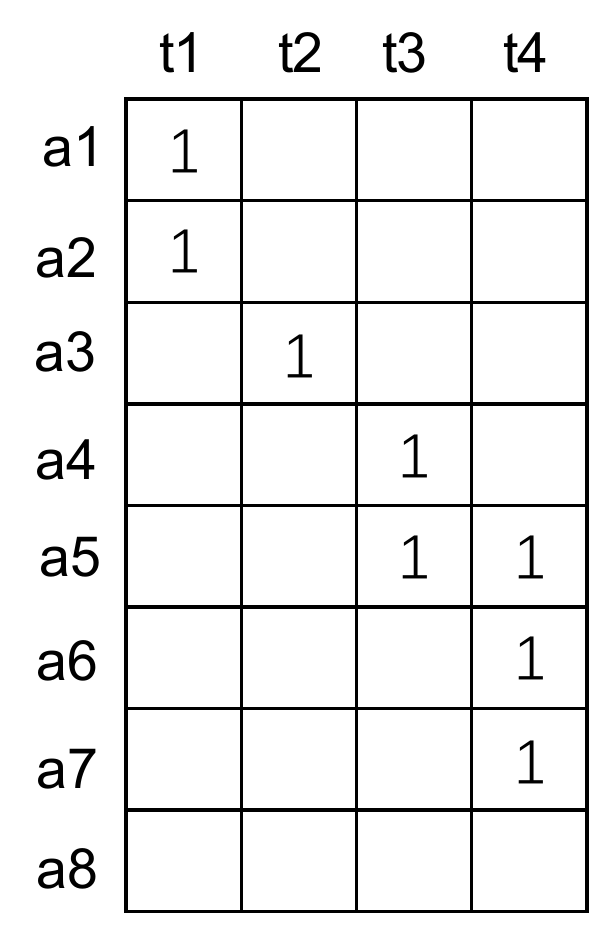}
	\caption{The $P$ matrix of Transaction Graph in Figure~\ref{fig:tx_graph}.}
	\label{fig:p_mat}
\end{figure}

Let $P \in \mathbb{R}^{|A|\times |T|}$ be the matrix denoting the input relationship between  each address $a_i$ and each transaction $t_j$. $P_{i,j} = 1$ if  $a_i$ is one of the input addresses of transaction $t_j$, otherwise, $P_{i,j} = 0$. Figure~\ref{fig:p_mat} shows The $P$ matrix of transaction graph in Figure~\ref{fig:tx_graph}.

\begin{figure}[h]
	\centering
	\includegraphics[width=0.7\linewidth]{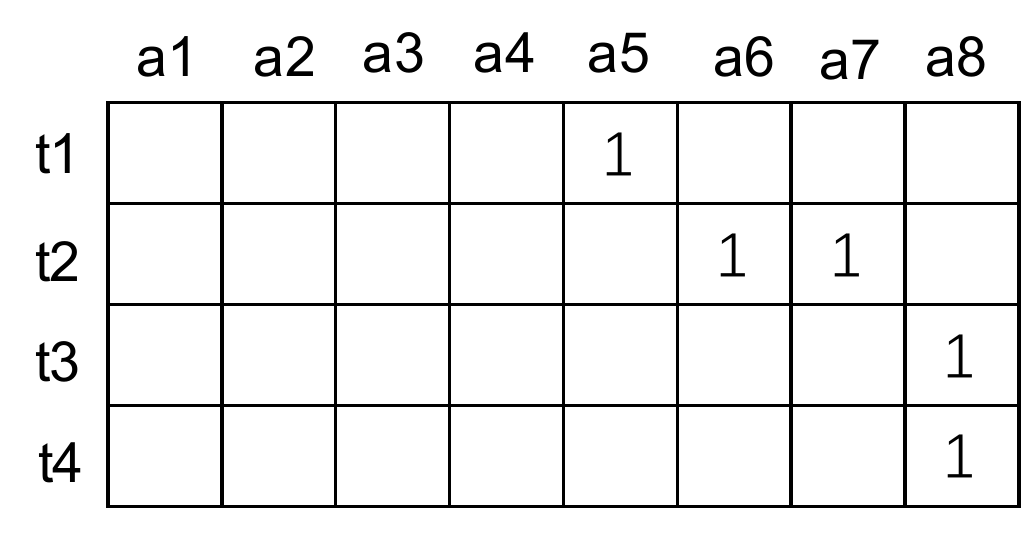}
	\caption{The $Q$ matrix of Transaction Graph in Figure~\ref{fig:tx_graph}.}
	\label{fig:q_mat}
\end{figure}

Then let $Q \in \mathbb{R}^{|T|\times |A|}$ be the matrix denoting the output relationship between each address $a_j$ and each transaction $t_i$. $Q_{i,j} = 1$ if  $a_j$ is one of the output addresses of transaction $t_i$, and $Q_{i,j} = 0$ otherwise. Figure~\ref{fig:q_mat} shows The $Q$ matrix of transaction graph in Figure~\ref{fig:tx_graph}.

For calculating the $k$ order occurrence matrix $OC^k$, we first need to derive the transition matrix $M\in \mathbb{R}^{|T|\times |A|}$ for the $k$ order transaction graph, which is derived through Equation~\ref{eq:1}. 

\begin{equation}
\label{eq:1}
M^k = H(QP)^{k-1}Q
\end{equation}

The entry of matrix $M^k$,  $M_{A_i, a_j}^k>0$ if there is a flow from transactions $t_i$, whose input addresses set is $A_i$, to address $a_j$, otherwise, $M_{A_i, a_j}^k=0$. In fact, we can easily understand that the  $M_{A_i, a_j}^k$ denotes how many possible path from transaction $t_i$ to address $a_j$ in the $k$ order transaction graph of transaction $t_i$.

Therefore $|A_i|$ is the number of input addresses of the  $k$ order transaction graph of $t_i$, and $\sum_{a_j \in A} \mathbb{I}\{M_{A_i, a_j}^k >0 \}$ is the number of output addresses of the  $k$ order transaction graph of $t_i$. $\mathbb{I}\{ *\} =1$ if the condition $*$ is satisfied, and  $\mathbb{I}\{ *\} =0$ otherwise.

Now each entry of $OC^k$, $OC_{(m,n)}^k$, can be calculated based on the $k$ order transition matrix $M^k$ through Equation~\ref{eq:2}.

\begin{equation}
\label{eq:2}
OC_{(m,n)}^k = \sum_{A_i} \mathbb{I}\{ |A_i|=m\  \& \sum_{a_j \in A} \mathbb{I}\{M_{A_i, a_j}^k >0 \} =n\},
\end{equation}
where $A_i$ is the set of input addresses of transaction $t_i$, namely $A_i = \{ a_k| (a_k,t_i) \in E\}$. 

\begin{figure}[h]
	\centering
	\includegraphics[width=0.9\linewidth]{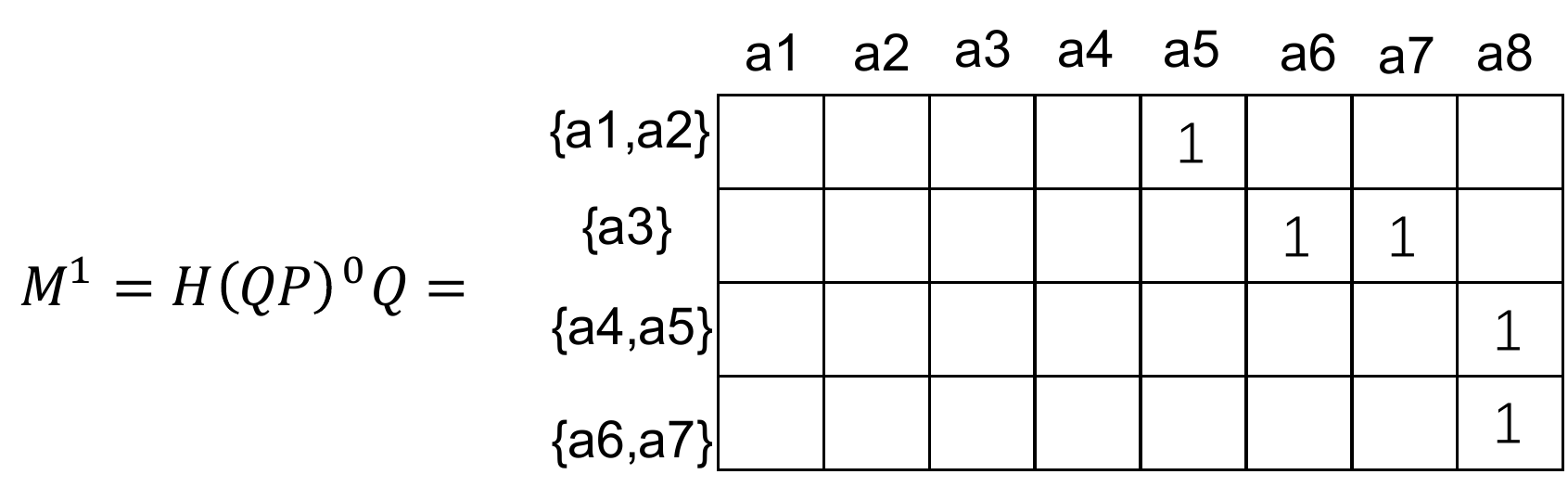}
	\caption{ $1$ Order Transition Matrix $M^1$ of Transaction Graph in Figure~\ref{fig:tx_graph}.}
	\label{fig:m1}
\end{figure}

For the simple example in Figure~\ref{fig:tx_graph}, if $k=1$, the transition matrix $M^1$ is illustrated in Figure~\ref{fig:m1}. Then the occurrence matrix $OC^1$ can be easily derived. %There are totally 4 $1$-order transaction subgraphs subject to 4 different  transactions.  

$OC^1_{(2,1)} =3$,  because $|\{a1,a2\}| = |\{a4,a5\}| = |\{a6,a7\}| = 2$, and  $\sum_{a_j \in A} \mathbb{I}\{M_{\{a1,a2\}, a_j}^1 >0 \} = \sum_{a_j \in A} \mathbb{I}\{M_{\{a4,a5\}, a_j}^1 >0 \}=\sum_{a_j \in A} \mathbb{I}\{M_{\{a6,a7\}, a_j}^1 >0 \}= 1$. 

$OC^1_{(1,2)} =1$, because $|\{a3\}| =1$, and $\sum_{a_j \in A} \mathbb{I}\{M_{\{a3\}, a_j}^1 >0 \} =2$. In addition, all the other entries of $OC^1$ is 0, since there is no other pattern for the $1$-order transactions graphs.

\begin{figure*}[h]
	\centering
	\includegraphics[width=0.8\linewidth]{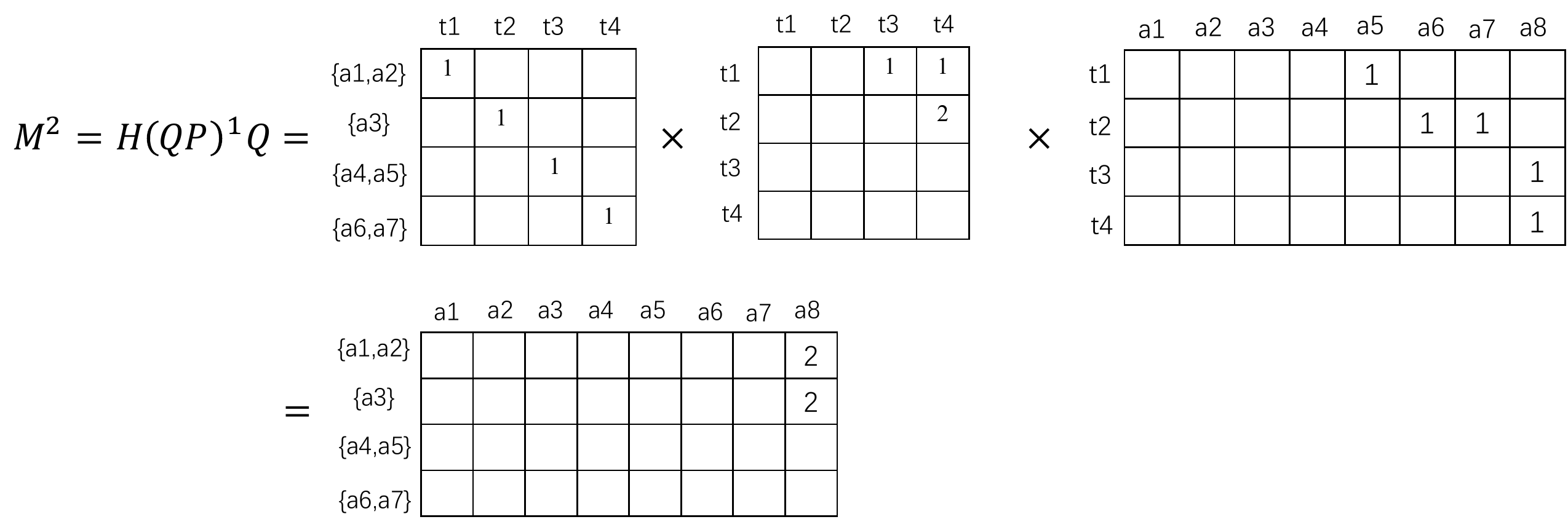}
	\caption{ $2$ Order Transition Matrix $M^2$ of Transaction Graph in Figure~\ref{fig:tx_graph}.}
	\label{fig:m2}
\end{figure*}

If $k=2$, the calculation of the transition matrix $M^2$ is illustrated in Figure~\ref{fig:m2}. Then the occurrence matrix $OC^2$ can be calculated as follows . $OC^2_{(2,1)} =1$, since $|\{a1,a2\}| =2$ and $\sum_{a_j \in A} \mathbb{I}\{M_{\{a1,a2\}, a_j}^1 >0 \}= 1$. $OC^2_{(1,1)} =1$, since $|\{a3\}| =1$ and $\sum_{a_j \in A} \mathbb{I}\{M_{\{a1\}, a_j}^1 >0 \}= 1$. All the other entries of $OC^2$ is 0. 

The dimension of occurrence matrices may be different for different $k$-order, and different transactions. However,  occurrence matrices of unified size are required for formating the feature vector of the same size, so that the features can be fed in machine learning based prediction models. According to literature~\cite{DBLP:conf/pakdd/AkcoraDGK18}, nearly 97.57\% transactions have the inputs and outputs sized no greater than 20. Therefore, for  the less than 3\% left transactions, whose number of inputs or outputs is greater than 20, we manually set number as 20. The $k$-order occurrence matrix $OC^k$ now can be defined as Equation~\ref{eq:OC}.

\begin{table*}
\begin{equation}
\label{eq:OC}
OC^k_{(m,n)}=\left\{
\begin{aligned}
\sum_{A_i} \mathbb{I}\{ |A_i|=m\  \& \sum_{a_j \in A} \mathbb{I}\{M_{A_i, a_j}^k >0 \} =n\} & , & m<20,n<20, \\
\sum_{A_i} \mathbb{I}\{ |A_i|\geq 20\  \& \sum_{a_j \in A} \mathbb{I}\{M_{A_i, a_j}^k >0 \} =n\} & , & m=20,n<20,\\
\sum_{A_i} \mathbb{I}\{ |A_i|=m\  \& \sum_{a_j \in A} \mathbb{I}\{M_{A_i, a_j}^k >0 \} \geq 20\} & , & m<20,n=20,\\
\sum_{A_i} \mathbb{I}\{ |A_i|\geq 20\  \& \sum_{a_j \in A} \mathbb{I}\{M_{A_i, a_j}^k >0 \} \geq 20\} & , & m=20,n=20.
\end{aligned}
\right.
\end{equation}
\end{table*}

\section{Experiments}
\label{sec:exp}

In this section, we present the evaluation of proposed transaction graph based blockchain feature and price prediction methods.

\subsection{Data preparation}
To conduct the bitcoin price prediction task, we collect the Bitcoin blockchain historical data and the bitcoin price historical data. The Bitcoin blockchain data is downloaded from Google Bigquery public dataset crypto\_bitcoin~\footnote{Dataset ID is bigquery-public-data: crypto\_bitcoin at https://cloud.google.com/bigquery} whose data is exported using bitcoin etl tool~\footnote{https://github.com/blockchain-etl/bitcoin-etl}. The bitcoin price data is collected from Coindesk~\footnote{https://www.coindesk.com/}. 

We select two history periods for the experiments. 

\begin{itemize}
	\item \textbf{Interval 1}: From August 19th, 2013 to July 19th, 2016. The timestamps are divided daily. This period contains 1065 days, the first 80\% days are for training and the left 20\% is for testing.
	\item \textbf{Interval 2}: From April 1st, 2013 to April 1st, 2017. The timestamps are divided daily. This period contains 1461 days, the first 70\% days are for training and the other 30\% is for testing.
\end{itemize}

\begin{figure}[h]
	\centering
	\includegraphics[width=0.9\linewidth]{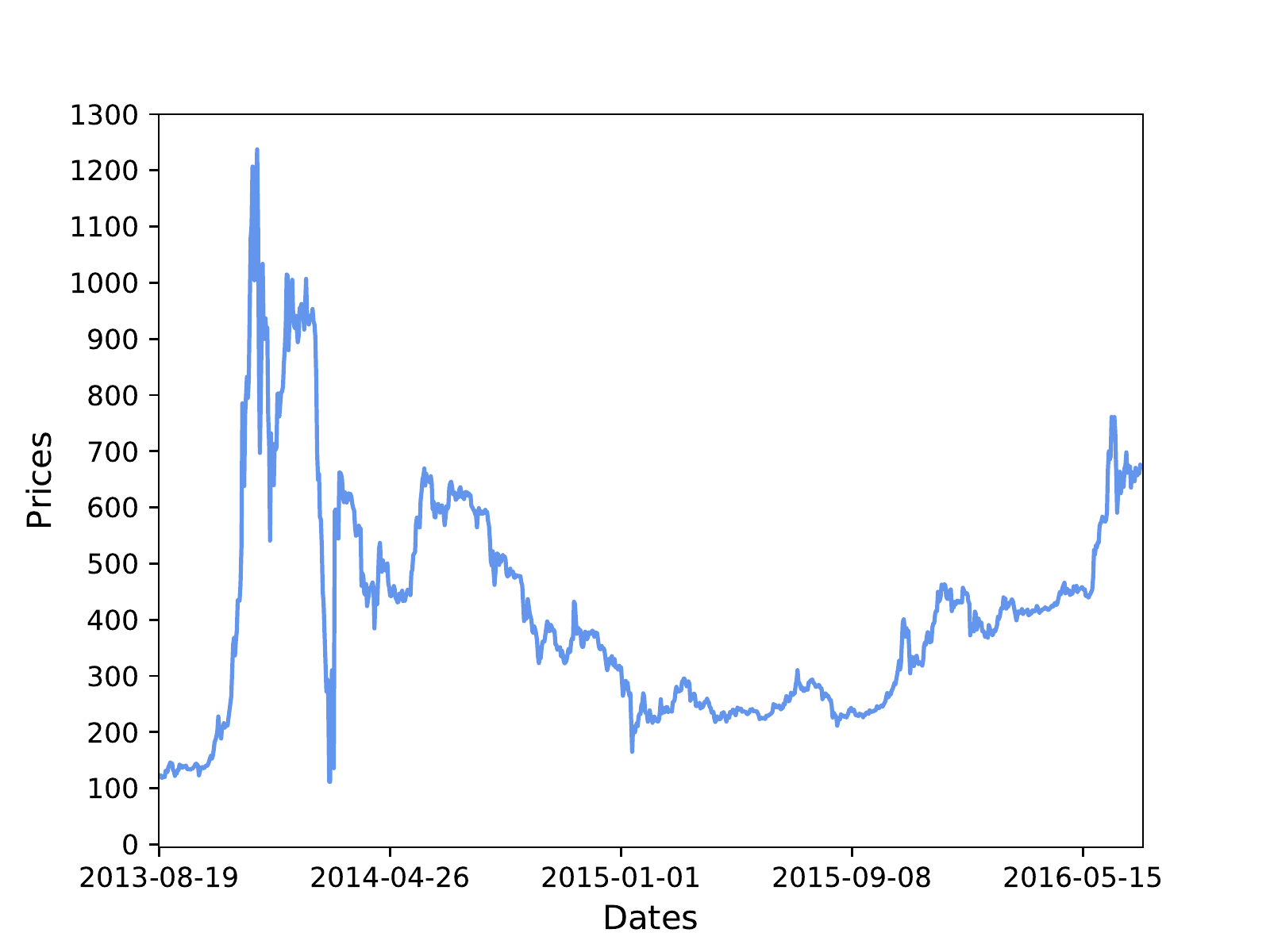}
	\caption{ Bitcoin Daily Closing Price of Interval 1}
	\label{fig:int_1}
\end{figure}

\begin{figure}[h]
	\centering
	\includegraphics[width=0.9\linewidth]{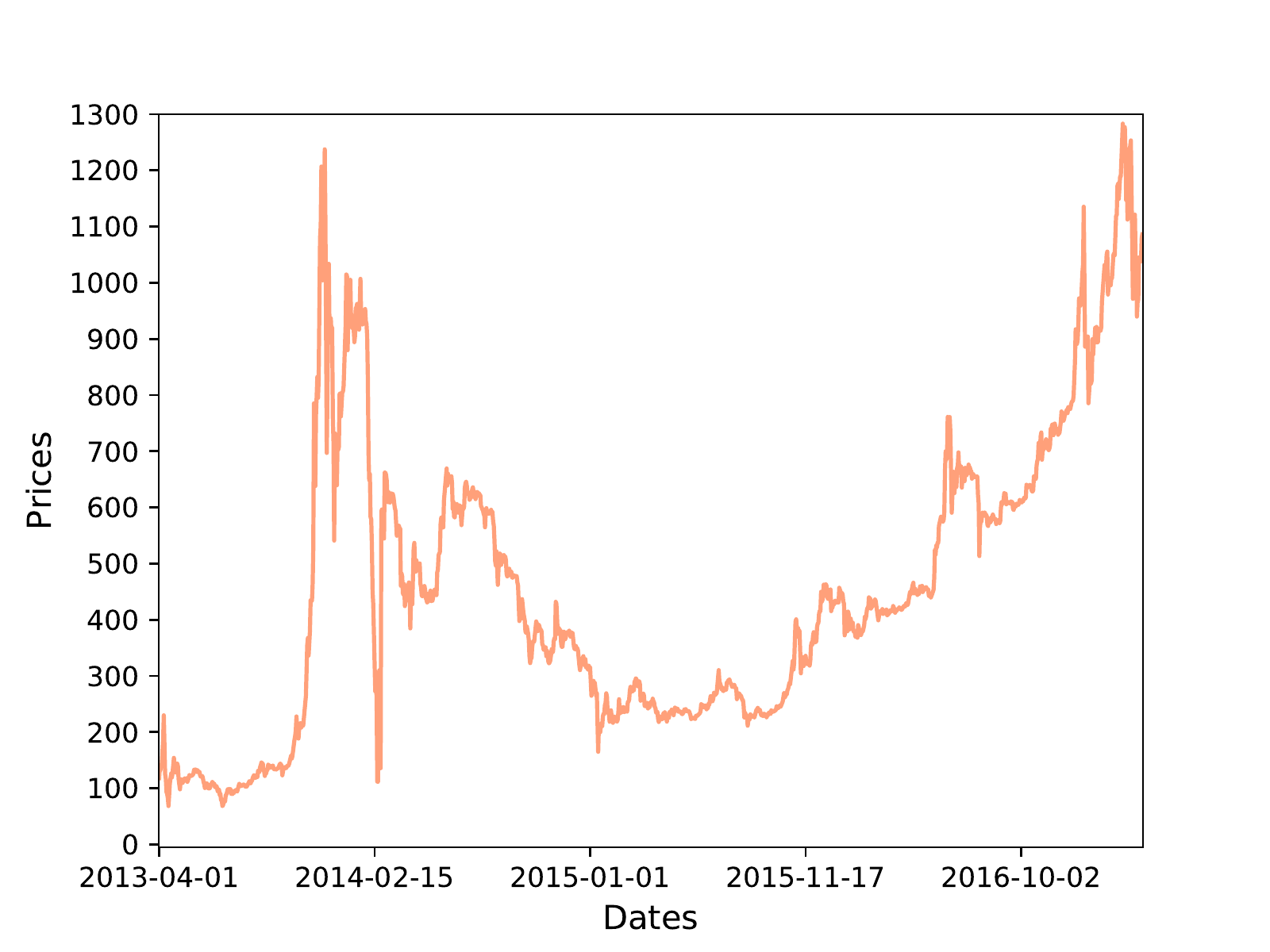}
	\caption{ Bitcoin Daily Closing Price of Interval 2}
	\label{fig:int_2}
\end{figure}

The \emph{interval 1} and \emph{interval 2} is identical to the datasets in literature~\cite{DBLP:journals/asc/MallquiF19}, which will be compared as a baseline in next sections. In this paper, we predict bitcoin daily closing price during the above periods. The bitcoin price of \emph{interval 1} and \emph{interval 2} is presented in Figure~\ref{fig:int_1} and Figure~\ref{fig:int_2}, respectively. It is possible to observe, that the bitcoin prices show a high volatility, which indicates that the nature of the bitcoin can hardly be intuitively discovered , therefore the features designed manually may be ineffective.

For the evaluation metric, we adopt Mean Absolute Percentage Error (MAPE) to show the error between predicted prices and real prices. The MAPE is formally defined as Equation~\ref{eq:mape}.
\begin{equation}
\label{eq:mape}
MAPE = \frac{1}{N} \sum_{i=1}^N \frac{|\hat{p_i}-p_i|}{p_i}
\end{equation}
where $\hat{p_i}$ is the predicted bitcoin price, while $p_i$ is the real price.

\subsection{Baselines and Comparison Results}
\subsubsection{Baselines}
We select Mallqui et. al. \cite{DBLP:journals/asc/MallquiF19} 's work as the baseline, where similar bitcoin price prediction task is studied. Mallqui et. al.  propose various attributes of Bitcoin Blockchain as features, including history price, volume of trades and even the finacial indicators from other financial area such as the Gold price and Nasdaq price. Mallqui et. al. utilize several machine learning methods to make prediction of bitcoin price based on the proposed features. 

Specifically, among the machine learning methods in \cite{DBLP:journals/asc/MallquiF19}, the SVM model shows the best prediction performance, therefore we adopt the SVM prediction model as the baseline prediction model, denoted as $Mallqui et. al. -SVM$.

\subsubsection{Results}
\begin{figure}[h]
	\centering
	\includegraphics[width=0.9\linewidth]{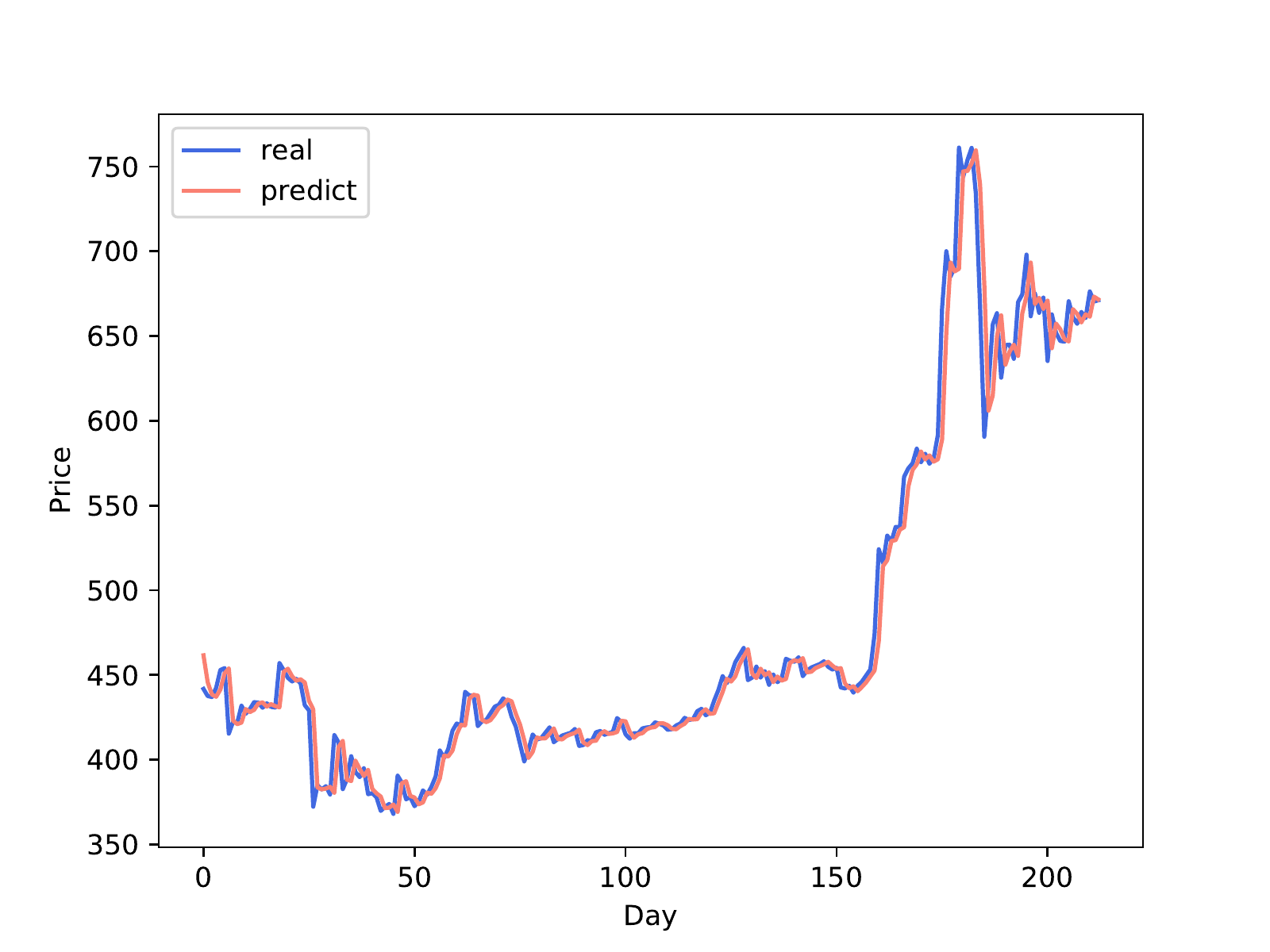}
	\caption{Price Prediction Results on \emph{interval1}}
	\label{fig:re_int_1}
\end{figure}

\begin{figure}[h]
	\centering
	\includegraphics[width=0.9\linewidth]{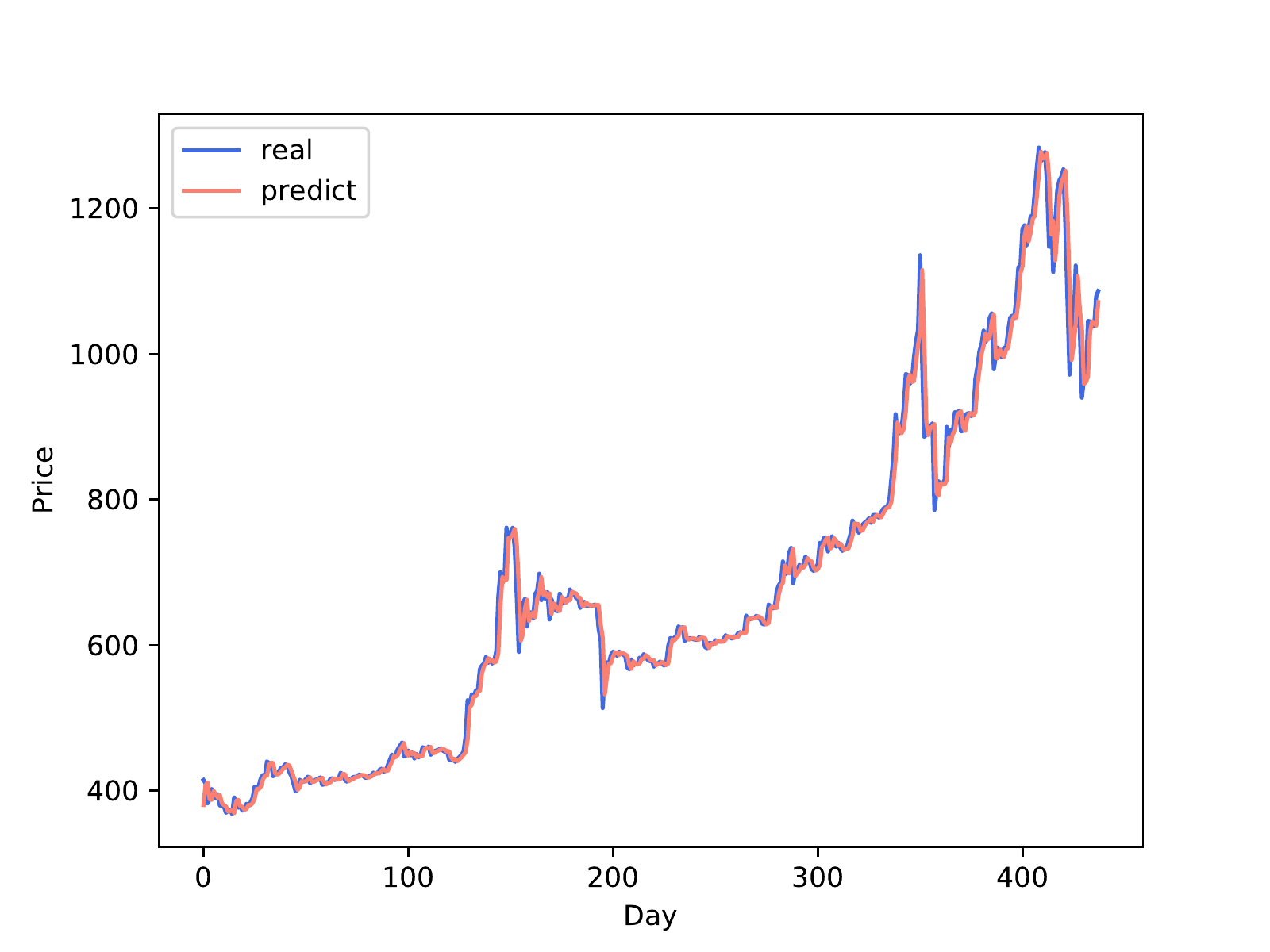}
	\caption{ Price Prediction Results on \emph{interval2}}
	\label{fig:re_int_2}
\end{figure}

To provide intuitive understanding of the price prediction results, we visualize the best results at \emph{interval1} and \emph{interval2} in Figure~\ref{fig:re_int_1} and Figure~\ref{fig:re_int_2} respectively.

The numerical results in terms of $MAPE$ is presented in Table~\ref{tab:re}. The results are obtained by the SVM version of our method and under $k=2$ and $r=0.8$. The results show that when $h=1$, that we only train the model for predict the next day, the results achieve lower MAPE than the baseline that also only trains one single model. Since both the baseline and our method here utilize the SVM model, the results can demonstrate the effectiveness of our proposed transaction graph based blockchain feature. 

Further, when we train more models for both $h=1$ and $h=2$, and integrate the results together as the final results, the performance can be even better. Therefore, the results demonstrate out proposed prediction method, that is to consider and combine the different latent features under different historical period,  can boost the accuracy of the prediction.

\begin{table}[h]
	\centering
	%\small
	\caption{MAPE of Baseline and Our Method on Two Time Periods}
	\label{tab:re}
	\begin{tabular}{|c|c|c|} \hline
		Methods$\backslash$ Periods & \emph{interval1} & \emph{interval2} \\ \hline
		$Mallqui et. al. -SVM$\cite{DBLP:journals/asc/MallquiF19} & 1.91\%& 1.81\% \\ \hline
		Our(h=1) & 1.75\% & 1.754\% \\ \hline
		Our(h=2) & 2.58\% &  2.590\% \\ \hline
		Our(integrated) & 1.69\%  &  1.751\% \\ \hline
	\end{tabular}
\end{table}

\subsection{Accuracy for future time $t'$}
\begin{figure}[htb]
	\centering
	\includegraphics[width=0.9\linewidth]{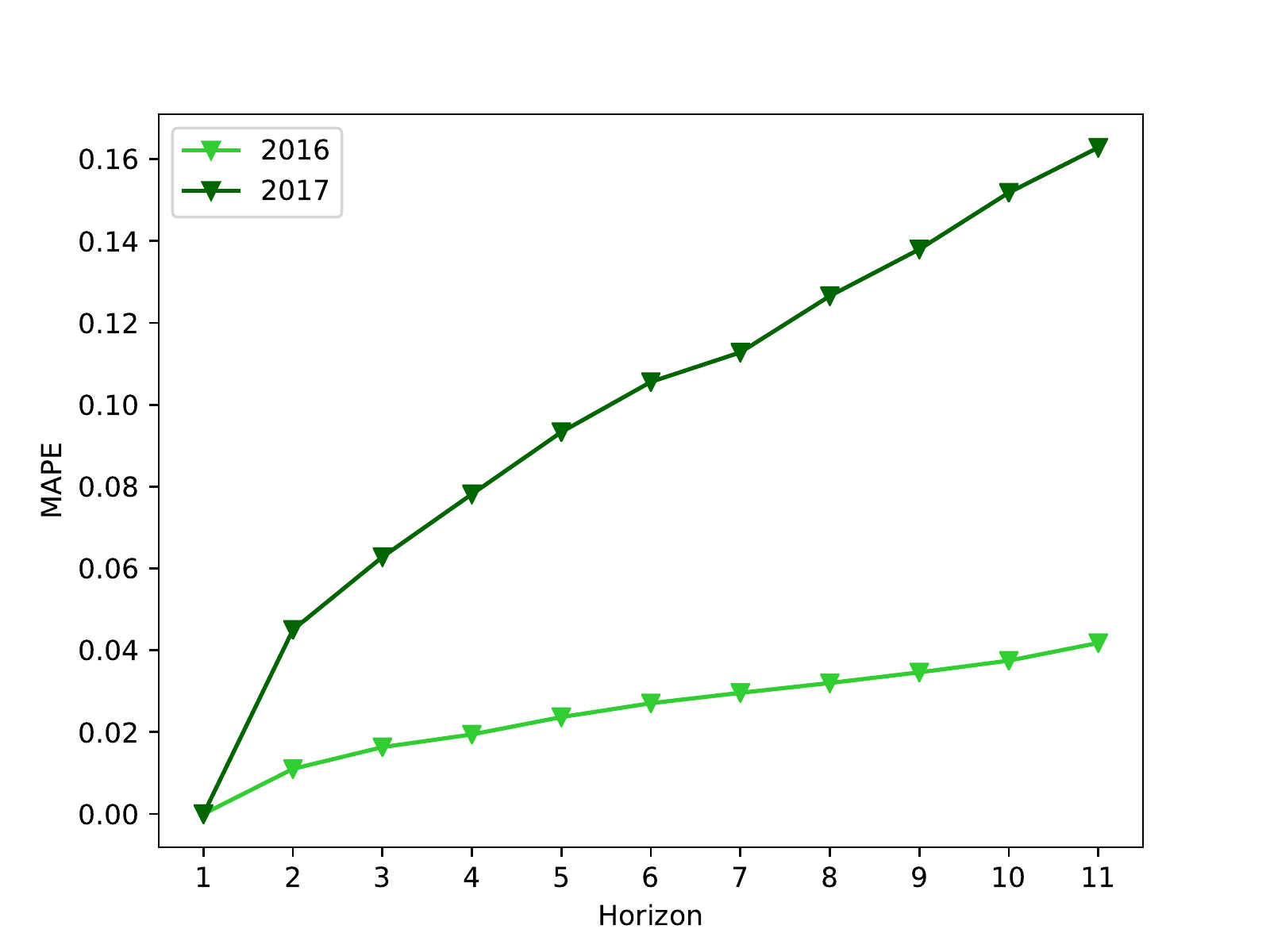}
	\caption{ MAPE of predicting price at different horizon on 2016 and 2017 Bitcoin market}
	\label{fig:diff_h}
\end{figure}

We study the accuracy for prediction of different future time stamps $t'$. Intuitively, further future is harder to predict, which indicates that the MAPE value will decrease with the increase of distance of future (also called horizon). We conduct experiments on 2016 and 2017 Bitcoin Blockchain respectively, and the results are illustrated in Figure~\ref{fig:diff_h}. We set $k=2$ and history window size as 1. 

The results is consistent with the intuition, that the MAPE keeps rising when we try to predict further future. In addition, the results demonstrate that the speed of growth differs with periods. The error rises faster in 2017 than 2016, which indicates that predicting the price in 2017 is much harder than it in 2016. This reflects the fact that the bitcoin prices fluctuate much strikingly in 2017 than 2016. 

\subsection{Influence of the length of history window}
\begin{figure}[hbt]
	\centering
	\includegraphics[width=0.9\linewidth]{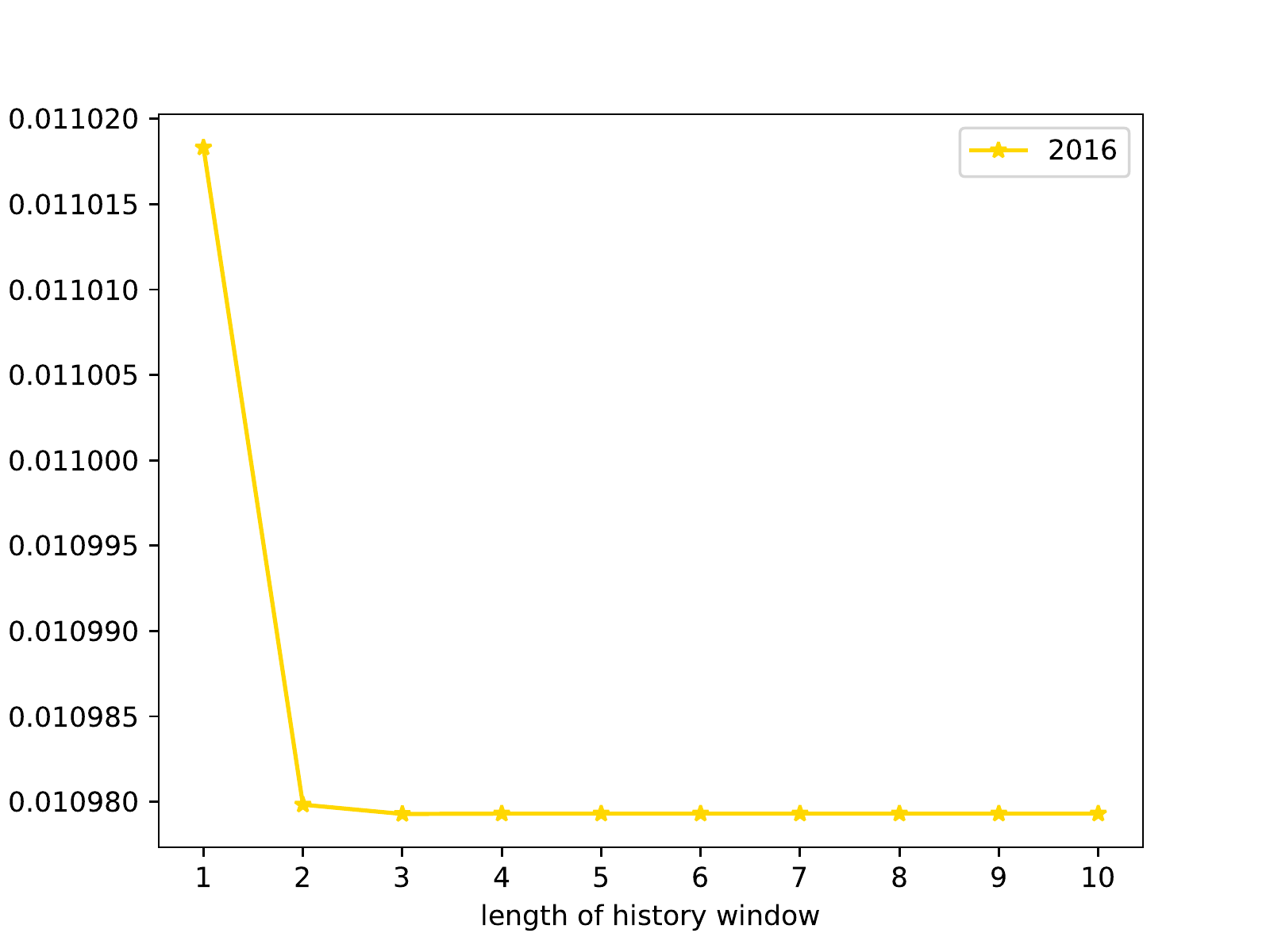}
	\caption{ MAPE of predicting price on 2016  Bitcoin market with different length of history window}
	\label{fig:his_2016}
\end{figure}

\begin{figure}[hbt]
	\centering
	\includegraphics[width=0.9\linewidth]{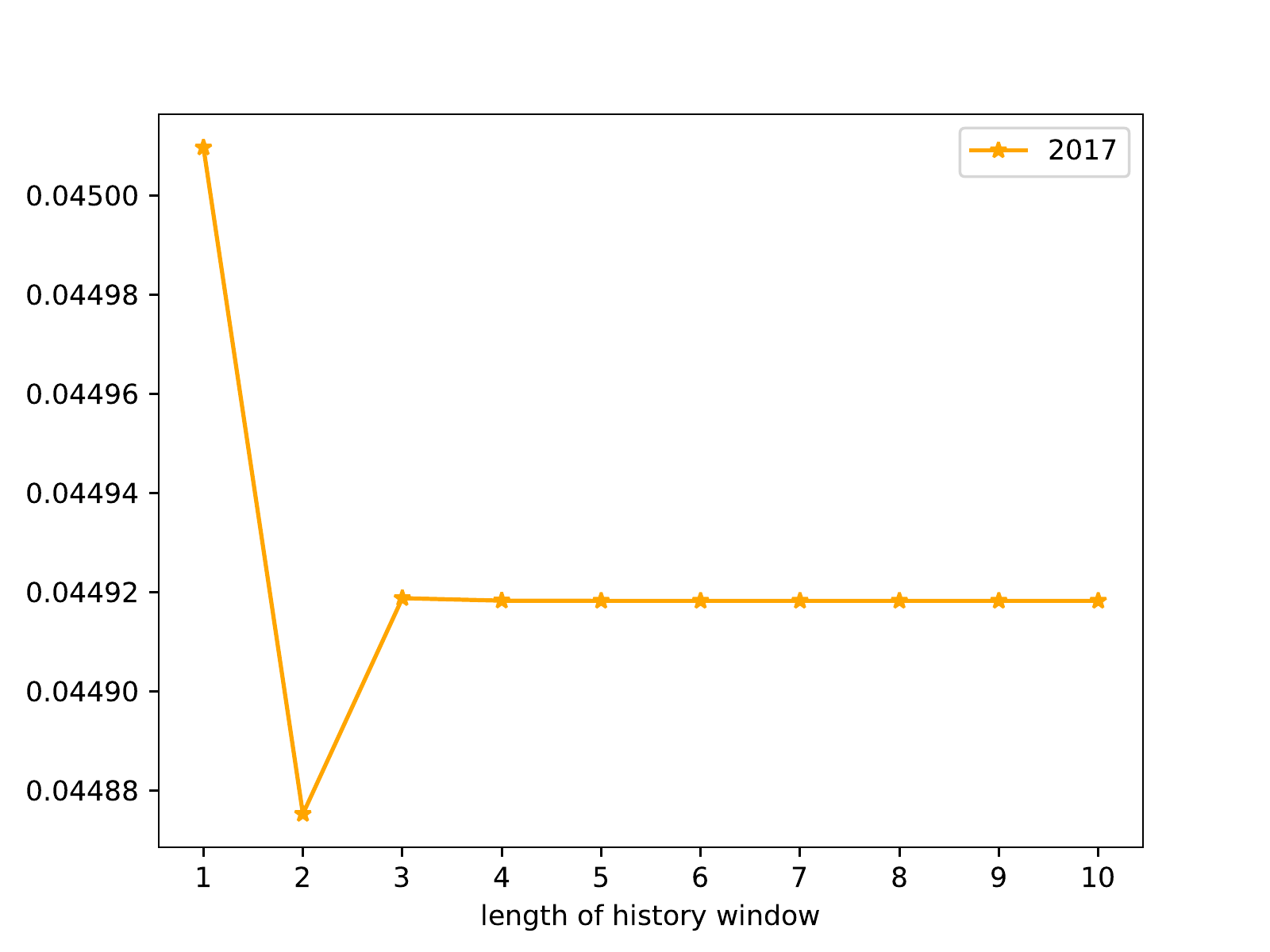}
	\caption{ MAPE of predicting price on 2017  Bitcoin market with different length of history window}
	\label{fig:his_2017}
\end{figure}
The length of history window size indicates how much historical information are utilized for the prediction. Since our proposed method trains models for different history window sizes and integrate the results as final results,  history window size can therefore impact the performance of whole framework. We compare the results obtained under different history window size in 2016 and 2017, and the results are illustrated in Figure~\ref{fig:his_2016} and Figure~\ref{fig:his_2017} respectively. We set $r=0.98$ and $r=0.92$ for 2016 and 2017 respectively, since this setting achieves best performance in our test.

The result in Figure~\ref{fig:his_2016} and Figure~\ref{fig:his_2017}  demonstrates that introducing more than one history time stamps (days in this paper) can significantly improve the prediction results. However, involving too much historical information may leading increase of computational consumption, since each length of history time stamps will be assigned one model to train the features. The results also shows that the performance of the framework will be relatively stable when the length of history window is greater or equal to 2. Therefore, in practice the framework will not consume too much to reach the best performance.

\section{Conclusions}
\label{sec:con}

In this paper, we proposed a transaction graph based machine learning method for bitcoin price prediction. The $k$-order transaction graphs of the transactions are proposed to reveal the transaction patterns in Bitcoin blockchain. The occurrence matrix is then defined to encode the patterns information and further be represented as the features of Bitcoin blockchain. We also provide mathematical formula for  iterative implementation of the features. Results of comparison experiments show that our proposed method outperforms the most recent state-of-art method, and demonstrate the effectiveness of automatically learning the transaction patterns from multiple blockchain history periods.


%\section*{Acknowledgment}

\bibliographystyle{IEEEtran}
\bibliography{ref}

\begin{thebibliography}{10}
\providecommand{\url}[1]{#1}
\csname url@samestyle\endcsname
\providecommand{\newblock}{\relax}
\providecommand{\bibinfo}[2]{#2}
\providecommand{\BIBentrySTDinterwordspacing}{\spaceskip=0pt\relax}
\providecommand{\BIBentryALTinterwordstretchfactor}{4}
\providecommand{\BIBentryALTinterwordspacing}{\spaceskip=\fontdimen2\font plus
\BIBentryALTinterwordstretchfactor\fontdimen3\font minus
  \fontdimen4\font\relax}
\providecommand{\BIBforeignlanguage}[2]{{%
\expandafter\ifx\csname l@#1\endcsname\relax
\typeout{** WARNING: IEEEtran.bst: No hyphenation pattern has been}%
\typeout{** loaded for the language `#1'. Using the pattern for}%
\typeout{** the default language instead.}%
\else
\language=\csname l@#1\endcsname
\fi
#2}}
\providecommand{\BIBdecl}{\relax}
\BIBdecl

\bibitem{Nakamoto2009BitcoinA}
S.~Nakamoto, ``Bitcoin : A peer-to-peer electronic cash system,'' 2009.

\bibitem{vassiliadis2017bitcoin}
S.~Vassiliadis, P.~Papadopoulos, M.~Rangoussi, T.~Konieczny, and J.~Gralewski,
  ``Bitcoin value analysis based on cross-correlations,'' \emph{Journal of
  Internet Banking and Commerce}, vol.~22, no.~S7, p.~1, 2017.

\bibitem{Aalborg2019WhatCE}
H.~A. Aalborg, P.~Moln{\'a}r, and J.~E. de~Vries, ``What can explain the price,
  volatility and trading volume of bitcoin?'' \emph{Finance Research Letters},
  vol.~29, pp. 255--265, 2019.

\bibitem{balcilar2017can}
M.~Balcilar, E.~Bouri, R.~Gupta, and D.~Roubaud, ``Can volume predict bitcoin
  returns and volatility? a quantiles-based approach,'' \emph{Economic
  Modelling}, vol.~64, pp. 74--81, 2017.

\bibitem{Yermack2013IsB}
D.~L. Yermack, ``﻿is bitcoin a real currency? an economic appraisal,''
  \emph{Economics of Innovation eJournal}, 2013.

\bibitem{DBLP:journals/jcam/ChenLS20}
\BIBentryALTinterwordspacing
Z.~Chen, C.~Li, and W.~Sun, ``Bitcoin price prediction using machine learning:
  An approach to sample dimension engineering,'' \emph{J. Comput. Appl. Math.},
  vol. 365, 2020. [Online]. Available:
  \url{https://doi.org/10.1016/j.cam.2019.112395}
\BIBentrySTDinterwordspacing

\bibitem{DBLP:conf/iscc/YaoXL19}
\BIBentryALTinterwordspacing
W.~Yao, K.~Xu, and Q.~Li, ``Exploring the influence of news articles on bitcoin
  price with machine learning,'' in \emph{2019 {IEEE} Symposium on Computers
  and Communications, {ISCC} 2019, Barcelona, Spain, June 29 - July 3,
  2019}.\hskip 1em plus 0.5em minus 0.4em\relax {IEEE}, 2019, pp. 1--6.
  [Online]. Available: \url{https://doi.org/10.1109/ISCC47284.2019.8969596}
\BIBentrySTDinterwordspacing

\bibitem{DBLP:conf/icdm/AbayAGKITT19}
\BIBentryALTinterwordspacing
N.~C. Abay, C.~G. Akcora, Y.~R. Gel, M.~Kantarcioglu, U.~D. Islambekov,
  Y.~Tian, and B.~M. Thuraisingham, ``Chainnet: Learning on blockchain graphs
  with topological features,'' in \emph{2019 {IEEE} International Conference on
  Data Mining, {ICDM} 2019, Beijing, China, November 8-11, 2019}, J.~Wang,
  K.~Shim, and X.~Wu, Eds.\hskip 1em plus 0.5em minus 0.4em\relax {IEEE}, 2019,
  pp. 946--951. [Online]. Available:
  \url{https://doi.org/10.1109/ICDM.2019.00105}
\BIBentrySTDinterwordspacing

\bibitem{DBLP:conf/pakdd/AkcoraDGK18}
\BIBentryALTinterwordspacing
C.~G. Akcora, A.~K. Dey, Y.~R. Gel, and M.~Kantarcioglu, ``Forecasting bitcoin
  price with graph chainlets,'' in \emph{Advances in Knowledge Discovery and
  Data Mining - 22nd Pacific-Asia Conference, {PAKDD} 2018, Melbourne, VIC,
  Australia, June 3-6, 2018, Proceedings, Part {III}}, ser. Lecture Notes in
  Computer Science, D.~Q. Phung, V.~S. Tseng, G.~I. Webb, B.~Ho, M.~Ganji, and
  L.~Rashidi, Eds., vol. 10939.\hskip 1em plus 0.5em minus 0.4em\relax
  Springer, 2018, pp. 765--776. [Online]. Available:
  \url{https://doi.org/10.1007/978-3-319-93040-4\_60}
\BIBentrySTDinterwordspacing

\bibitem{DBLP:journals/asc/MallquiF19}
\BIBentryALTinterwordspacing
D.~C.~A. Mallqui and R.~A.~S. Fernandes, ``Predicting the direction, maximum,
  minimum and closing prices of daily bitcoin exchange rate using machine
  learning techniques,'' \emph{Appl. Soft Comput.}, vol.~75, pp. 596--606,
  2019. [Online]. Available: \url{https://doi.org/10.1016/j.asoc.2018.11.038}
\BIBentrySTDinterwordspacing

\bibitem{DBLP:conf/www/CerdaR19}
\BIBentryALTinterwordspacing
G.~C. Cerda and J.~L. Reutter, ``Bitcoin price prediction through opinion
  mining,'' in \emph{Companion of The 2019 World Wide Web Conference, {WWW}
  2019, San Francisco, CA, USA, May 13-17, 2019}, S.~Amer{-}Yahia, M.~Mahdian,
  A.~Goel, G.~Houben, K.~Lerman, J.~J. McAuley, R.~Baeza{-}Yates, and L.~Zia,
  Eds.\hskip 1em plus 0.5em minus 0.4em\relax {ACM}, 2019, pp. 755--762.
  [Online]. Available: \url{https://doi.org/10.1145/3308560.3316454}
\BIBentrySTDinterwordspacing

\bibitem{DBLP:conf/dsaa/MaesaMR16}
\BIBentryALTinterwordspacing
D.~D.~F. Maesa, A.~Marino, and L.~Ricci, ``Uncovering the bitcoin blockchain:
  An analysis of the full users graph,'' in \emph{2016 {IEEE} International
  Conference on Data Science and Advanced Analytics, {DSAA} 2016, Montreal, QC,
  Canada, October 17-19, 2016}.\hskip 1em plus 0.5em minus 0.4em\relax {IEEE},
  2016, pp. 537--546. [Online]. Available:
  \url{https://doi.org/10.1109/DSAA.2016.52}
\BIBentrySTDinterwordspacing

\bibitem{kristoufek2013bitcoin}
L.~Kristoufek, ``Bitcoin meets google trends and wikipedia: Quantifying the
  relationship between phenomena of the internet era,'' \emph{Scientific
  reports}, vol.~3, no.~1, pp. 1--7, 2013.

\bibitem{DBLP:conf/bigdataconf/BalfagihK19}
\BIBentryALTinterwordspacing
A.~M. Balfagih and V.~Keselj, ``Evaluating sentiment c1assifiers for bitcoin
  tweets in price prediction task,'' in \emph{2019 {IEEE} International
  Conference on Big Data (Big Data), Los Angeles, CA, USA, December 9-12,
  2019}.\hskip 1em plus 0.5em minus 0.4em\relax {IEEE}, 2019, pp. 5499--5506.
  [Online]. Available: \url{https://doi.org/10.1109/BigData47090.2019.9006140}
\BIBentrySTDinterwordspacing

\bibitem{DBLP:journals/ijecommerce/PolasikPWKL15}
\BIBentryALTinterwordspacing
M.~Polasik, A.~I. Piotrowska, T.~P. Wisniewski, R.~Kotkowski, and G.~Lightfoot,
  ``Price fluctuations and the use of bitcoin: An empirical inquiry,''
  \emph{Int. J. Electron. Commer.}, vol.~20, no.~1, pp. 9--49, 2015. [Online].
  Available: \url{https://doi.org/10.1080/10864415.2016.1061413}
\BIBentrySTDinterwordspacing

\bibitem{DBLP:conf/ic3/MittalDSP19}
\BIBentryALTinterwordspacing
A.~Mittal, V.~Dhiman, A.~Singh, and C.~Prakash, ``Short-term bitcoin price
  fluctuation prediction using social media and web search data,'' in
  \emph{2019 Twelfth International Conference on Contemporary Computing, {IC3}
  2019, Noida, India, August 8-10, 2019}.\hskip 1em plus 0.5em minus
  0.4em\relax {IEEE}, 2019, pp. 1--6. [Online]. Available:
  \url{https://doi.org/10.1109/IC3.2019.8844899}
\BIBentrySTDinterwordspacing

\bibitem{DBLP:conf/sigir/BurnieY19}
\BIBentryALTinterwordspacing
A.~Burnie and E.~Yilmaz, ``An analysis of the change in discussions on social
  media with bitcoin price,'' in \emph{Proceedings of the 42nd International
  {ACM} {SIGIR} Conference on Research and Development in Information
  Retrieval, {SIGIR} 2019, Paris, France, July 21-25, 2019}, B.~Piwowarski,
  M.~Chevalier, {\'{E}}.~Gaussier, Y.~Maarek, J.~Nie, and F.~Scholer,
  Eds.\hskip 1em plus 0.5em minus 0.4em\relax {ACM}, 2019, pp. 889--892.
  [Online]. Available: \url{https://doi.org/10.1145/3331184.3331304}
\BIBentrySTDinterwordspacing

\bibitem{DBLP:conf/mcis/GeorgoulaPBSG15}
\BIBentryALTinterwordspacing
I.~Georgoula, D.~Pournarakis, C.~Bilanakos, D.~N. Sotiropoulos, and G.~M.
  Giaglis, ``Using time-series and sentiment analysis to detect the
  determinants of bitcoin prices,'' in \emph{9th Mediterranean Conference on
  Information Systems, {MCIS} 2015, Samos, Greece, October 2-5, 2015.
  Proceedings}.\hskip 1em plus 0.5em minus 0.4em\relax AISeL, 2015, p.~20.
  [Online]. Available: \url{http://aisel.aisnet.org/mcis2015/20}
\BIBentrySTDinterwordspacing

\bibitem{Ciaian16}
\BIBentryALTinterwordspacing
P.~Ciaian, M.~Rajcaniova, and d’Artis Kancs, ``The economics of bitcoin price
  formation,'' \emph{Applied Economics}, vol.~48, no.~19, pp. 1799--1815, 2016.
  [Online]. Available: \url{https://doi.org/10.1080/00036846.2015.1109038}
\BIBentrySTDinterwordspacing

\bibitem{brandvold2015price}
M.~Brandvold, P.~Moln{\'a}r, K.~Vagstad, and O.~C.~A. Valstad, ``Price
  discovery on bitcoin exchanges,'' \emph{Journal of International Financial
  Markets, Institutions and Money}, vol.~36, pp. 18--35, 2015.

\bibitem{DBLP:conf/ic3/AggarwalGGG19}
\BIBentryALTinterwordspacing
A.~Aggarwal, I.~Gupta, N.~Garg, and A.~Goel, ``Deep learning approach to
  determine the impact of socio economic factors on bitcoin price prediction,''
  in \emph{2019 Twelfth International Conference on Contemporary Computing,
  {IC3} 2019, Noida, India, August 8-10, 2019}.\hskip 1em plus 0.5em minus
  0.4em\relax {IEEE}, 2019, pp. 1--5. [Online]. Available:
  \url{https://doi.org/10.1109/IC3.2019.8844928}
\BIBentrySTDinterwordspacing

\bibitem{DBLP:journals/iepol/PietersV17}
\BIBentryALTinterwordspacing
G.~Pieters and S.~Vivanco, ``Financial regulations and price inconsistencies
  across bitcoin markets,'' \emph{Inf. Econ. Policy}, vol.~39, pp. 1--14, 2017.
  [Online]. Available: \url{https://doi.org/10.1016/j.infoecopol.2017.02.002}
\BIBentrySTDinterwordspacing

\bibitem{kristoufek2015main}
L.~Kristoufek, ``What are the main drivers of the bitcoin price? evidence from
  wavelet coherence analysis,'' \emph{PloS one}, vol.~10, no.~4, p. e0123923,
  2015.

\bibitem{bouoiyour2016drives}
J.~Bouoiyour, R.~Selmi, A.~K. Tiwari, O.~R. Olayeni \emph{et~al.}, ``What
  drives bitcoin price,'' \emph{Economics Bulletin}, vol.~36, no.~2, pp.
  843--850, 2016.

\bibitem{DBLP:journals/ijcse/ChenZMWZY20}
\BIBentryALTinterwordspacing
W.~Chen, Z.~Zheng, M.~Ma, J.~Wu, Y.~Zhou, and J.~Yao, ``Dependence structure
  between bitcoin price and its influence factors,'' \emph{{IJCSE}}, vol.~21,
  no.~3, pp. 334--345, 2020. [Online]. Available:
  \url{https://doi.org/10.1504/IJCSE.2020.106058}
\BIBentrySTDinterwordspacing

\bibitem{DBLP:conf/educon/YogeshwaranKM19}
\BIBentryALTinterwordspacing
S.~Yogeshwaran, M.~J. Kaur, and P.~Maheshwari, ``Project based learning:
  Predicting bitcoin prices using deep learning,'' in \emph{{IEEE} Global
  Engineering Education Conference, {EDUCON} 2019, Dubai, United Arab Emirates,
  April 8-11, 2019}, A.~K. Ashmawy and S.~Schreiter, Eds.\hskip 1em plus 0.5em
  minus 0.4em\relax {IEEE}, 2019, pp. 1449--1454. [Online]. Available:
  \url{https://doi.org/10.1109/EDUCON.2019.8725091}
\BIBentrySTDinterwordspacing

\bibitem{DBLP:conf/icnc/SinW17}
\BIBentryALTinterwordspacing
E.~Sin and L.~Wang, ``Bitcoin price prediction using ensembles of neural
  networks,'' in \emph{13th International Conference on Natural Computation,
  Fuzzy Systems and Knowledge Discovery, {ICNC-FSKD} 2017, Guilin, China, July
  29-31, 2017}, Y.~Liu, L.~Zhao, G.~Cai, G.~Xiao, K.~Li, and L.~Wang,
  Eds.\hskip 1em plus 0.5em minus 0.4em\relax {IEEE}, 2017, pp. 666--671.
  [Online]. Available: \url{https://doi.org/10.1109/FSKD.2017.8393351}
\BIBentrySTDinterwordspacing

\bibitem{DBLP:conf/besc/FelizardoOHC19}
\BIBentryALTinterwordspacing
L.~Felizardo, R.~Oliveira, E.~Del{-}Moral{-}Hernandez, and F.~Cozman,
  ``Comparative study of bitcoin price prediction using wavenets, recurrent
  neural networks and other machine learning methods,'' in \emph{6th
  International Conference on Behavioral, Economic and Socio-Cultural
  Computing, {BESC} 2019, Beijing, China, October 28-30, 2019}.\hskip 1em plus
  0.5em minus 0.4em\relax {IEEE}, 2019, pp. 1--6. [Online]. Available:
  \url{https://doi.org/10.1109/BESC48373.2019.8963009}
\BIBentrySTDinterwordspacing

\bibitem{DBLP:conf/ispacs/ChenCLHLW19}
\BIBentryALTinterwordspacing
C.~Chen, J.~Chang, F.~Lin, J.~Hung, C.~Lin, and Y.~Wang, ``Comparison of
  forcasting ability between backpropagation network and {ARIMA} in the
  prediction of bitcoin price,'' in \emph{2019 International Symposium on
  Intelligent Signal Processing and Communication Systems, {ISPACS} 2019,
  Taipei, Taiwan, December 3-6, 2019}.\hskip 1em plus 0.5em minus 0.4em\relax
  {IEEE}, 2019, pp. 1--2. [Online]. Available:
  \url{https://doi.org/10.1109/ISPACS48206.2019.8986297}
\BIBentrySTDinterwordspacing

\bibitem{DBLP:conf/icdm/WuLML18}
\BIBentryALTinterwordspacing
C.~Wu, C.~Lu, Y.~Ma, and R.~Lu, ``A new forecasting framework for bitcoin price
  with {LSTM},'' in \emph{2018 {IEEE} International Conference on Data Mining
  Workshops, {ICDM} Workshops, Singapore, Singapore, November 17-20, 2018},
  H.~Tong, Z.~J. Li, F.~Zhu, and J.~Yu, Eds.\hskip 1em plus 0.5em minus
  0.4em\relax {IEEE}, 2018, pp. 168--175. [Online]. Available:
  \url{https://doi.org/10.1109/ICDMW.2018.00032}
\BIBentrySTDinterwordspacing

\bibitem{DBLP:conf/etfa/HashishFAFD19}
\BIBentryALTinterwordspacing
I.~A. Hashish, F.~Forni, G.~Andreotti, T.~Facchinetti, and S.~Darjani, ``A
  hybrid model for bitcoin prices prediction using hidden markov models and
  optimized {LSTM} networks,'' in \emph{24th {IEEE} International Conference on
  Emerging Technologies and Factory Automation, {ETFA} 2019, Zaragoza, Spain,
  September 10-13, 2019}.\hskip 1em plus 0.5em minus 0.4em\relax {IEEE}, 2019,
  pp. 721--728. [Online]. Available:
  \url{https://doi.org/10.1109/ETFA.2019.8869094}
\BIBentrySTDinterwordspacing

\bibitem{DBLP:conf/fdse/NguyenL19}
\BIBentryALTinterwordspacing
D.~Nguyen and H.~Le, ``Predicting the price of bitcoin using hybrid {ARIMA} and
  machine learning,'' in \emph{Future Data and Security Engineering - 6th
  International Conference, {FDSE} 2019, Nha Trang City, Vietnam, November
  27-29, 2019, Proceedings}, ser. Lecture Notes in Computer Science, T.~K.
  Dang, J.~K{\"{u}}ng, M.~Takizawa, and S.~H. Bui, Eds., vol. 11814.\hskip 1em
  plus 0.5em minus 0.4em\relax Springer, 2019, pp. 696--704. [Online].
  Available: \url{https://doi.org/10.1007/978-3-030-35653-8\_49}
\BIBentrySTDinterwordspacing

\end{thebibliography}

\begin{IEEEbiographynophoto}{Xiao Li}
received his
B.S. and M.S degree in Software Engineering from Dalian University of Technology, China in
2016 and 2019, respectively. He is currently pursuing the
Ph.D. degree with the Department of Computer
Science, University of Texas at Dallas, Richardson,
TX, USA.
His current research interests include  data mining and Blockchain.
\end{IEEEbiographynophoto}

\begin{IEEEbiographynophoto}{Weili Wu}
 received the Ph.D. and M.S. degrees from
the Department of Computer Science, University of
Minnesota, Minneapolis, MN, USA, in 2002 and
1998, respectively.

She is currently a Full Professor with the Department of Computer Science, The University of
Texas at Dallas, Richardson, TX, USA. Her current research interests include data communication,
data management, the design and analysis of algorithms for optimization problems that occur in wireless networking environments, and various database
systems.
\end{IEEEbiographynophoto}

\end{document}